\documentclass[nofootinbib, amsmath,amssymb, aps,]{revtex4}

\usepackage{hyperref}

\usepackage{graphicx}
\usepackage{dcolumn}
\usepackage{bm}

\usepackage{booktabs}
\usepackage{topcapt}

\usepackage{xcolor}

\usepackage{slashed} 

 \usepackage{ulem}
 \usepackage{cancel}

\begin{document}
\title{QED photon-fermion vertex from its Dyson-Schwinger equation in 4D: The full vertex, the transverse form factors and the perturbative solution}
\author{Orlando Oliveira}

\affiliation{CFisUC, Departament of Physics, University of Coimbra, 3004-516 Coimbra, Portugal}

\begin{abstract}
We investigate the Dyson-Schwinger equation for the photon-fermion one-particle irreducible vertex in QED in linear covariant gauges. 
The longitudinal component of this vertex is described using the Ball-Chiu basis, while its transverse part is expressed with the K{\i}z{\i}lersu-Reenders-Pennington basis. 
Combining the vertex Ward-Takahashi identity with the vertex equation, we derive a set of exact, nonlinear integral equations governing the transverse vertex. 
These equations hold for any linear covariant gauge and must be solved self-consistently. We discuss several approximations to the exact equations, generalizing 
results previously obtained at the perturbative one-loop level. Various kinematical configurations are also examined. Furthermore, we compute the perturbative solution 
of the transverse vertex Dyson-Schwinger equations for all transverse form factors and derive their perturbative asymptotic expressions.
\end{abstract}

\maketitle
\tableofcontents

\section{Introduction and Motivation}

The traditional approach to solve Quantum Electrodynamics (QED) relies on perturbation theory (PT), which provides the standard diagrammatic representation of the solution, 
and is also at the heart of our understanding of Quantum Field Theory (QFT). Although important, PT cannot  tell the full story as there are phenomena, such us
chiral symmetry breaking, whose explanation lies beyond  this approach. This is particularly true when the coupling is large enough, challenging the applicability of PT itself. 

For QED the expansion parameter is the fine-structure constant  that has the value $\alpha = 7.297 352 5693(11) \times 10^{-3}$ \cite{ParticleDataGroup:2024cfk}.
Its smallness supports the use of perturbative methods. However, one might question whether the perturbative expansion converges or  define
an asymptotic series, see e.g. \cite{ZinnJustin} and references therein. For large couplings, perturbation theory becomes unreliable and  new phenomena emerge. 
For example, at sufficiently large couplings the theory predicts a phase transition, that can be associated with dynamical mass generation, see e.g.
\cite{Fukuda:1976zb,Miransky:1984ef,Miransky:1985wzx,Gusynin:1986fu,Miransky:1989qc,Oliveira:2024tne} and references therein.
The study of QED at large couplings is not a purely academic problem. It  offers insights into the solutions and phenomena expected to occur 
within Quantum Chromodynamics (QCD), see e.g. \cite{Grozin:2007zz,Huber:2018ned,Eichmann:2022zxn,Ferreira:2025anh}.
Exploring either QED at strong coupling or QCD requires non-perturbative methods.

The infinite tower of Dyson-Schwinger equations (DSE), which relate all the Green functions of the theory
provide a way to solve any QFT, regardless  of the coupling strength. The DSE 
allow  to go beyond the traditional perturbative approach, offering a powerful alternative to analyze and solve quantum field theories.

This work aims  to study QED in a general linear covariant gauge, defined by a gauge fixing parameter $\xi$, using the framework of the
DSE. The goal is to combine simultaneously the integral equations for the two point functions, i.e. the fermion and photon propagators,
and the equation for the photon-fermion vertex, to gather information on the vertex from first principles, and to explore other gauges than the Landau gauge.  
Hopefully, this approach will deliver insights to help understanding the DSE for non-Abelian gauge theories, that include the Standard 
Model of Particle Physics or QCD. 
A precise understanding of the  photon-fermion vertex has implications extending far beyond QED.
In the Standard Model of Particle Physics, a detailed knowledge of this vertex is essential to characterize other
vector boson couplings, both to disentangle the contributions from different interactions and to probe potential signatures of Physics Beyond the Standard Model.

Another first-principles approach to QFT are Monte Carlo lattice simulations. These rely on the Euclidean formulation of the theories and are not
suitable to be used in its Minkowski spacetime version. 
The lattice simulations requires a careful control of finite-volume and finite-spacing effects, together with the use of realistic physical parameters. 
For full QCD, the vertex longitudinal form factors where computed, in the soft gluon limit, recently in \cite{Kizilersu:2021jen}. 
Previous lattice estimations of the form factors using the quenched approximation can be found in \cite{Kizilersu:2006et,Skullerud:2003qu,Skullerud:2002ge}
.
For QED, to the best knowledge of the author, the only lattice calculation of the photon-fermion vertex was reported in the proceeding \cite{Sternbeck:2019twy}. 
While these results provide benchmarks, 
a proper separation of strong interaction effects from pure QED contributions must be performed before any meaningful comparison can be made.

The investigation of QED through DSE presents numerous challenges, which must be addressed by building upon the extensive existing literature on DSE 
applications to gauge theories. While a comprehensive citation of all relevant works would be impractical, we have endeavored to maintain a focused bibliography and apologize for any 
omissions.

We will not reproduce the derivation of the integral equations here. The complete derivation, along with notational conventions, which we follow, can be found in \cite{Oliveira:2022bar}. 
The DSE formalism itself is well-documented in the literature. For a general discussion of DSE and their application to QED, we refer readers to \cite{Roberts:1994dr} among other 
sources. The application of DSE to QCD is discussed in e.g. \cite{Bardeen:1990im,Alkofer:2000wg,Fischer:2006ub,RichardWilliams2007,Braun:2010qs,Antipin:2012kc,Gao:2021wun,Ding:2022ows,Binosi:2022djx,Papavassiliou:2022wrb,Alkofer:2023lrl,Aguilar:2023mdv,Ferreira:2023fva} (see also the references therein).

The vertex Dyson-Schwinger equation, together with the vertex Ward-Takahashi identity (WTI), provide a complete framework for calculating all twelve form factors required to fully 
describe this one-particle irreducible Green function. As is well established, the longitudinal form factors are  determined by the WTI and depend solely on the fermion propagator 
functions. 
The computation of the transverse form factors builds on these results and, within the approach adopted herein,
rely on the vertex integral equation. 

From the vertex DSE 
a set of exact nonlinear equations, that must be solved self-consistently, is derived.
These equations apply to any general kinematical situation. However, some special kinematical situations will be discussed to help
on the understanding of the vertex. 
In particular, the case with on-shell external fermions, leading to the on-shell vertex, from where one can read, for example,
 the fermion anomalous magnetic moments is investigated. 
 This anomalous magnetic moment will be expressed in terms of the fundamental form factors associated with the photon-fermion vertex.
 
 In order to grasp the properties of the transverse form factors, approximations to the exact equations, that go beyond the usual perturbative approach, are discussed 
 and, within  the framework considered, it is shown that, in the chiral limit, the number of form factors describing the vertex is smaller than twelve. This finding generalizes
 the conclusions coming from the one-loop perturbative solution of QED for a general covariant gauge. 
 Moreover, our analysis facilitates the study of the various form factors for the full class of linear covariant gauges. 
 Being a first step towards  a complete computation of the QED photon-fermion vertex, the current study ends looking at the perturbative solution, for all
 transverse form factors, of the equations derived. Expressions for their asymptotic behaviour are  provided.
 Whenever possible, the solutions for the vertex are compared to the outcome of a one-loop perturbative calculation.

This work is organized as follows. In Sec. \ref{Sec:Intro} the basic definitions are introduced, also to set the notation. In \ref{Sec:Renormalizacao} the renormalization of
QED is revisited and the integral equations to be studied are given in terms of physical quantities. A perturbative description of the renormalized equations, together
with a brief discussion of infrared divergences, is given at the end of the section. The basis used to describe the vertex is introduced in Sec. \ref{Sec:vertice-tensor}, together with the solutions of the vertex Ward-Takahashi identity
for its longitudinal components. In the current work,  the Ball-Chiu set of operators \cite{Ball:1980ay} is used for the longitudinal vertex,  
while for the orthogonal components we rely on the operators introduced by K{\i}z{\i}lersu-Reenders-Pennington \cite{Kizilersu:1995iz}.
In Sec. \ref{Sec:PropEqFF} we refer to the structure of the integral equations using the full vertex, that are given in the appendices.
In Sec. \ref{SecVertexMain} the tensor decomposition of the vertex is discussed, together with the consideration of special kinematical configurations.
In Sec. \ref{Sec:OnShell} an on-shell vertex is defined and its form factors are given in terms of the form factors that are associated with the basis used to describe the vertex.
In \ref{SecTransverseFF} the computation of the eight transverse form factors is discussed based on vertex Dyson-Schwinger integral equation. The exact
equations derived have to be solved self-consistently as they  give the transverse form factor in terms of the full vertex itself, that is a function of all the longitudinal and transverse
form factors.
In Sec. \ref{Sec:softphotonlimit} an approximation scheme  to the vertex equation, inspired in the setup built from perturbation theory, is discussed. 
In Sec. \ref{Sec:PertFFactors}, a perturbative analysis of the Dyson-Schwinger vertex equation is provided, including the discussion of the
asymptotic behaviour for all non-vanishing transverse form factors.
Finally, in Sec. \ref{Sec:Summary} we summarize and conclude.

\section{Definitions and Integral Equations \label{Sec:Intro}}

Let us introduce the notation to be used through out this work. Herein, all expressions are given in Minkowski spacetime.
For  linear covariant gauges, the photon propagator in momentum space reads
\begin{equation}
    D_{\mu\nu}(k) = - \left( g_{\mu\nu} - \frac{k_\mu k_\nu}{k^2} \right) \, D(k^2) - \frac{\xi}{k^2} \, \frac{k_\mu k_\nu}{k^2} 
    = - P^\perp_{\mu\nu} (k) \, D(k^2) - \frac{\xi}{k^2} \, P^L_{\mu \nu}(k)  \ .
\end{equation}
The Lorentz invariant scalar function $D(k^2)$ will be referred also as the photon propagator. The inverse of the fermion propagator is given by
\begin{equation}
 S^{-1}(p) = A(p^2) \, \slashed{p} - B(p^2) + i \, \epsilon \ ,
\end{equation}
where $A(p^2)$ and $B(p^2)$ are Lorentz scalar functions and the limit $\epsilon \rightarrow 0^+$ is to be taken at the end of the calculations.
In the following, unless clearly stated, the $i \epsilon$ term will be omitted from now on.

The photon-fermion one-particle irreducible Green function (1PI), also named photon-fermion vertex,
 will be written as $\Gamma^\mu (p, -p-k; k)$, where $p$ is the incoming fermion
momentum, $p+k$ is the outgoing fermion momentum and $k$ the incoming photon momentum. This Green function has to comply with
the QED gauge symmetry that relates its longitudinal part, relative to the photon momentum, to the fermion propagator via a Ward-Takahashi identity (WTI).
The Ward-Takahashi vertex identity determines \cite{Ball:1980ay} (not-uniquely) the longitudinal part of $\Gamma^\mu$. 
We postpone the discussion on the vertex to later sections.

The bare DSE to be consider are 
\begin{equation}
  S^{-1} (p)  = \left(  \slashed{p} - m \right)
     - \, i \, g^2 \, \int \frac{d^4 k}{( 2 \, \pi )^4} ~   D_{\mu\nu} (k) ~  \Big[ \gamma^\mu ~ S(p - k) ~   {\Gamma}^\nu (p-k, - p; k)  \Big] \ ,
     \label{DSE-Fermion}
\end{equation}
for the fermion propagator, and 
\begin{eqnarray}
  \frac{1}{D(k^2)}  = 
   k^2 
  - i \, \frac{g^2}{3} \,  \int \frac{d^4 p}{(2 \, \pi)^4} ~ \text{Tr} \Big[ \gamma_\mu \, S(p) \, {\Gamma}^\mu(p, -p + k; -k ) \, S(p - k ) \Big] \ ,
     \label{Eq:DSE-Photon}
\end{eqnarray}
for the photon propagator.
The bare Dyson-Schwinger equation for the photon-fermion vertex reads
\begin{eqnarray}
& & 
   {\Gamma}^\mu (p, \, -p -k; \, k)   =  \gamma^\mu  ~ + ~ i \, g^2 \, \int \frac{d^4q}{(2 \, \pi)^4} ~D_{\zeta\zeta^\prime}(q)  ~ \gamma^\zeta ~ S(p-q)
   \nonumber \\
& & 
\qquad
   \Bigg\{ ~ 
                                 \, {\Gamma}^\mu ( p - q, \, -p -k + q; \, k) \,
                                S(p+k-q) \, {\Gamma}^{\zeta^\prime} (p +k-q, \, -p-k; q) 
~ + ~
                                 \, {\Gamma}^{\zeta^\prime\mu} (p - q, \, -p-k; \, q , \, k) 
                                \Bigg\}  \ .
                                \label{Eq:DSE-vertex}
\end{eqnarray}
The DSE for the vertex requires, besides the propagators and $\Gamma^\mu$ itself, the one-particle irreducible two-photon-two-fermion Green function
$\Gamma^{\mu\nu}$.  
Gauge symmetry constrains non-trivially these two Green functions with $\Gamma^\mu$ and $\Gamma^{\mu\nu}$ having to
comply with WTI's. The vertex WTI has been discussed long ago \cite{Ball:1980ay}. The solution of the WTI for $\Gamma^{\mu\nu}$,
for general kinematics, can be found in \cite{Oliveira:2022bar}.
The solutions of the WTI determine, non-uniquely, the longitudinal components of the vertices; see the discussion in
\cite{Roberts:1994dr}.
The 1PI function $\Gamma^{\mu\nu}$ is the solution of its own DSE, that calls for higher order Green functions.
An approximation to the DSE for $\Gamma^{\mu\nu}$ can be found in \cite{Oliveira:2022bar} but, given its complexity, it will not be considered in this work and, when
building solutions of the above DSE, the contribution of $\Gamma^{\mu\nu}$ will be ignored. This is a drastic approximation that can be justified
within a perturbative solution of the theory, given that, to lowest order in the coupling constant, $\Gamma^{\mu\nu}$ vanish. In perturbation theory, to lowest
order in the coupling constant, this 1PI Green function appears at one-loop level and is of order $\mathcal{O}(g^4)$.

\section{Renormalized QED Equations\label{Sec:Renormalizacao}}

The renormalization constants $Z_i$ relate bare and physical quantities are defined as
\begin{eqnarray}
& &
   A_\mu  = Z^{\frac{1}{2}}_3 \, A^{(phys)}_\mu   , \quad
   \psi  = Z^{\frac{1}{2}}_2 \, \psi^{(phys)} , \quad
   g = \frac{Z_1}{Z_2 \, Z^{\frac{1}{2}}_3}      g^{(phys)}    ,  
\quad
   m = \frac{Z_0}{Z_2} \, m^{(phys)} \quad\mbox{and}\quad
  \xi = Z_3 \,  \xi^{(phys)}  ,
\end{eqnarray}
and allow to rewrite the DSE for the corresponding physical quantities. For QED the vertex WTI requires $Z_1 = Z_2$, see e.g. \cite{Roberts:1994dr,Oliveira:2022bar}, and
the renormalized gap equation (\ref{DSE-Fermion}) reads
\begin{eqnarray}
  S^{-1} (p)  & = &  Z_2 \,   \slashed{p} - Z_0 \, m 
     - \, i \, g^2  \, Z_2 \, \int \frac{d^4 k}{( 2 \, \pi )^4} ~   D_{\mu\nu} (k) ~  \Big[ \gamma^\mu ~ S(p - k) ~   {\Gamma}^\nu (p-k, - p; k)  \Big]    \nonumber  \\
     & = & 
     Z_2 \,   \slashed{p} - Z_0 \, m 
     - \, i \,   g^2 \, Z_2 \, \Sigma(p) \ ,
     \label{DSER-gap}
\end{eqnarray}
the renormalized photon gap equation (\ref{Eq:DSE-Photon}) is
\begin{eqnarray}
  \frac{1}{D(k^2)}  & = &
  Z_3 \,  k^2 
  - i \, \frac{g^2}{3} \,  Z_2 \,  \int \frac{d^4 p}{(2 \, \pi)^4} ~ \text{Tr} \Big[ \gamma_\mu \, S(p) \, {\Gamma}^\mu(p, -p + k; -k ) \, S(p - k ) \Big] 
  \nonumber \\
  & = & Z_3 \,  k^2  \left( 1  - i \,  g^2 \,  \frac{Z_2}{Z_3} \,   \Pi(k^2) \right) \nonumber \\
     \label{DSER-photon}
\end{eqnarray}
and the equation for the Green function (\ref{Eq:DSE-vertex}) is given by
\begin{eqnarray}
& & 
   {\Gamma}^\mu (p, \, -p -k; \, k)  ~  =  ~  Z_2 \, \gamma^\mu  ~ + ~ i \, g^2 \,  Z_2 \, \int \frac{d^4q}{(2 \, \pi)^4} ~D_{\zeta\zeta^\prime}(q) \nonumber \\
   & & \hspace{1cm}
    \Bigg\{
                                \gamma^\zeta \, S(p-q) \, {\Gamma}^\mu ( p - q, \, -p -k + q; \, k) \,
                                S(p+k-q) \, {\Gamma}^{\zeta^\prime} (p +k-q, \, -p-k; q) 
     \nonumber \\
    & & \hspace{8cm} 
                 ~ + ~
                                \gamma^\zeta \, S(p-q) \, {\Gamma}^{\zeta^\prime\mu} (p - q, \, -p-k; \, q , \, k) 
                                \Bigg\}  \, .
                                \label{DSER-vertex}
\end{eqnarray}
In the renormalized Eqs (\ref{DSER-gap}) to (\ref{DSER-vertex}) the index $(phys)$ was omitted to simplify the notation. 
The functions $\Sigma(p)$ and $\Pi(k^2)$ are, respectively, the fermion and the photon physical self-energies.
The renormalization constants can be computed requiring that
\begin{equation}
A(\mu^2_F) = 1, \qquad B(\mu^2_F) = m \qquad\mbox{ and }\qquad D(\mu^2_B) = \frac{1}{\mu^2_B}  \ ,
  \label{Renormalization-Conditions}
\end{equation}
where $\mu_F$ and $\mu_B$ are the renormalization mass scales for the fermion and boson fields.
Writing, as usual, the fermion self-energy as $\Sigma(p) = \Sigma_v(p^2) \, \slashed{p} + \Sigma_s(p^2)$ then 
\begin{eqnarray}
    Z_2 = {1 \over 1 - i \, g^2 \, \Sigma_v (\mu_F^2)} \, ,  
    \qquad
    Z_0 = 1  - i \,  {Z_2 \over m } g^2 \, \Sigma_s (\mu_F^2) 
    \qquad\mbox{ and }\qquad
    Z_3 = 1 + \, i \, Z_2 \, g^2 \,  \Pi( \mu^2_B)  \ . 
    \label{Eq:RenZ3}
\end{eqnarray}
In QED,  the combination $g^2 \, D(k^2)$ is independent of the renormalization scale, 
see e.g. \cite{Roberts:1994dr,Oliveira:2022bar} and references therein, and can be used to define an effective charge. The lowest order perturbative solution for
$D(k^2)$ is such that this effective charge diverges for small $k^2$ and vanish at large momentum scales. This effective charge is nothing but the Coulomb
potential if one assumes that there is no energy exchange.

\subsection{On the perturbative analysis of the DSE and the photon self-energy \label{Sec:PertSolPhoton}}

Let us assume that the coupling constant $g$ is small enough in such away that the theory can be solved using perturbation theory.
Then, inserting the expressions for $Z_i$ in the renormalized DSE, after some algebra, the integral equations can be approximate by
\begin{eqnarray}
A(p^2) & = & \frac{1 ~ - ~ i \, g^2 \, \Sigma_v(p^2)}{1 ~ - ~ i \, g^2 \, \Sigma_v(\mu^2_F)}  
               \approx 1 ~ - ~ i \, g^2 \, \Big( \Sigma_v(p^2)  ~ - ~  \Sigma_v(\mu^2_F)  \Big)  ~ + ~  \mathcal{O}(g^4) \ ,  \label{A-DSE-renor} \\
B(p^2) & = & m ~ + ~ i \, g^2 \, \frac{ \Sigma_s(p^2) - \Sigma_s(\mu^2_F)}{1 ~ - ~ i \, g^2 \, \Sigma_v(\mu^2_F)} 
               \approx m ~ + ~ i \, g^2 \, \Big(  \Sigma_s(p^2) - \Sigma_s(\mu^2_F) \Big) ~ + ~  \mathcal{O}(g^4)  \ , \label{B-DSE-renor} \\
\frac{1}{D(k^2)} & = & k^2 \left( 1 ~ - ~ i \, g^2 \,  \frac{ \Pi(k^2) - \Pi(\mu^2_B)}{1 ~ - ~ i \, g^2 \, \Sigma_v(\mu^2_F)}  \right) 
                \approx k^2  \bigg(  1 ~ - ~ i \, g^2 \,  \Big(  {\Pi}(k^2) - {\Pi}(\mu^2_B) \Big) \bigg) ~ + ~  \mathcal{O}(g^4)
\ . \label{D-DSE-renor} 
\end{eqnarray}
These subtractions  make the theory UV finite. Calling for dimensional regularization to perform the momentum integrations,
for example, the photon self-energy is in first order perturbation theory, in the chiral limit where $m = 0$, given by
\begin{eqnarray}
\frac{1}{D(k^2)} & = &  k^2  \Bigg( ~  1 ~ - ~ \frac{\alpha}{3 \, \pi}  \, N_\chi \,\ln \frac{k^2}{\mu^2_B } ~ \Bigg) \ , \label{D-DSE-renor-pert-DimReg-Chiral}  
\end{eqnarray}
where $N_\chi$ is the number of chiral fermions and $\alpha = g^2 / 4 \, \pi$.
In the chiral limit, first order pertubation theory predicts a pole for the photon propagator at 
$k^2 = 0$ together with a pole at
$k^2 = \mu^2_B \, \exp \left( 3 \, \pi / \alpha \, N_\chi \right)$, that moves to zero or to infinity as $\mu^2_B \rightarrow 0$ or
$\mu^2_B \rightarrow + \infty$, respectively. In the limit $\mu^2_B \rightarrow 0$, this later pole no longer shows up.
The pole goes away from $k^2 = 0$ as the number of chiral fermions $N_\chi$ increases.
On the other hand,
for $k^2 = 0$ and for a finite fermion mass, independently of the renormalization scale $\mu_B$, the perturbative result reads
\begin{eqnarray}
\frac{1}{D(0)} & = & - \, \frac{2 \, \alpha }{3 \, \pi} \, N_\chi  \, m^2                            \ ,
\end{eqnarray}
that appears as a photon mass term that grows linearly with the mass  of the fermion within the loop. 
This effective photon mass vanishes in the limit of massless fermions, if one assumes that the theory behaves smoothly in the limit of a vanishing
fermion mass.

The infrared problems  in QED are understood as being associated with having a massless photon. Gauge symmetry does not allow for a mass term at the level
of the Lagrangian density. However, perturbation theory requires an IR regulator as the theory is plagued with divergences that are associated with the
low momentum regions. For the DSE, possible IR divergences can be investigated studying the renormalized equations
at zero momentum, as considered in the last paragraph. 
This discussion becomes clear after the consideration of a tensor basis for the photon-fermion vertex as is done in Sec. \ref{Sec:vertice-tensor}. 
In what concerns the IR properties of the propagators, the fermion gap shows no naive IR divergences. However, to have a finite and IR safe equation for
the photon equation the replacement  of  ${\Pi}(k^2)$ by $ {\Pi}(k^2) -  {\Pi}(0)$ should be considered. 
The redefinition of the photon self-energy is equivalent to renormalize the theory at zero momentum, implying  a $Z_3 = 1$. 

The IR behaviour of the vertex DSE is more involved, when compared to the IR properties of the two-point functions.
This equation requires the knowledge of the two-photon-two-fermion vertex when one of the photon momenta vanishes.  
For this kinematics, the WTI for the two-photon-two-fermion vertex was solved in \cite{Oliveira:2022bar}. However, this solution of the WTI
only gives the longitudinal component of $\Gamma^{\mu\nu}$, that is described in terms of the fermion propagator form factors $A(p^2)$ and $B(p^2)$,
providing no information on its remaining components. 
In general, the tensor decomposition of $\Gamma^{\mu\nu}$ demands a basis with a large number of operators \cite{Bardeen:1968ebo,Perrottet:1973qw,Tarrach:1975tu}, 
making the comprehension of their IR properties 
and, therefore, the vertex behaviour at lower momenta outside perturbation theory difficult to analyse.
Anyway, by choosing to solve the integral equations within a perturbative approach, then $\Gamma^{\mu\nu}$ can be ignored on a first step, and the usual
perturbative divergences show up.
For the discussion of infrared problems within QED see e.g. 
\cite{Bloch:1937pw,Yennie:1961ad,Kinoshita:1962ur,Lee:1964is,Gamboa:2025dry}
and references therein.

\section{The  photon-fermion vertex \label{Sec:vertice-tensor}}

The photon-fermion Green function $\Gamma^\mu$ appears in all the DSE considered so far. Its description calls for twelve form factors \cite{Ball:1980ay}
and it is common to write $\Gamma^\mu$ as a sum of a longitudinal $\Gamma_L$ and a transverse $\Gamma_T$ part, relative to the photon 
momentum, i.e. 
\begin{equation}
 \Gamma^\mu (p_2, \, p_1; \, p_3)  = \Gamma^\mu_L (p_2, \, p_1; \, p_3) +  \Gamma^\mu_T (p_2, \, p_1; \, p_3) \ ,
\end{equation}
where $p_2$ is the incoming fermion momentum, $-p_1$ is the outgoing fermion momentum and $p_3$ is the incoming photon momentum.
Given that all the momenta are incoming they verify the relation $p_1 + p_2 + p_3 = 0$. From the definition of the transverse 
vertex it follows that
\begin{equation}
{ p_{3} } _{\mu } \,  \Gamma^\mu_T (p_2, \, p_1; \, p_3)  = 0 \ .
 \end{equation}
The analysis of the vertex DSE, see Eq. (\ref{Eq:DSE-vertex}), becomes
easier after the introduction of a tensor basis describing its longitudinal and transverse components. Let
us write
 \begin{eqnarray}
 \Gamma_{L \, \mu} (p_2, \, p_1; p_3) & = & \sum^4_{i=1} \lambda_i (p^2_1, \, p^2_2, \, p^2_3) \, L^{(i)}_\mu (p_1, \, p_2, \,  p_3) \ ,  
 \label{Eq:photon-fermion_vertex-longitudinal}  \\
  \Gamma_{T \, \mu} (p_2, \, p_1; p_3) & = &  \sum^8_{i=1} \tau_i (p^2_1, \, p^2_2, \, p^2_3) \, T^{(i)}_\mu (p_1, \, p_2, \,  p_3) \ ,
 \label{Eq:photon-fermion_vertex-transverse}
\end{eqnarray}
where $L^{(i)}_\mu$ and $T^{(i)}_\mu$ are the set of tensor operators that define the basis for the vertex, and $\lambda_i$ and $\tau_i$
are Lorentz scalar form factors. Note the different ordering of the momenta in the l.h.s. and r.h.s in Eqs. (\ref{Eq:photon-fermion_vertex-longitudinal})
and  (\ref{Eq:photon-fermion_vertex-transverse}). For the longitudinal part of the vertex we take the Ball-Chiu longitudinal set of operators \cite{Ball:1980ay} 
that are given by 
\begin{eqnarray}
L^{(1)}_\mu (p_1, \, p_2, \,  p_3) & = & \gamma_\mu \ ,  \label{TensorBasis-L1} \\
L^{(2)}_\mu (p_1, \, p_2, \,  p_3) & = & \big( \slashed{p}_1 - \slashed{p}_2 \big) \big( p_{1} - p_{2} \big)_\mu \ , \\
L^{(3)}_\mu (p_1, \, p_2, \,  p_3) & = & \big( p_{1} - p_{2} \big)_\mu \ , \\
L^{(4)}_\mu (p_1, \, p_2, \,  p_3) & = & \sigma_{\mu\nu} \big( p_{1} - p_{2} \big)^\nu \ ,
\label{TensorBasis-Long} 
\end{eqnarray}
while for the orthogonal part of the vertex we rely on the K{\i}z{\i}lersu-Reenders-Pennington basis \cite{Kizilersu:1995iz} 
\begin{eqnarray}
T^{(1)}_\mu (p_1, \, p_2, \,  p_3) & = & p_{1 \, _\mu} \big( p_2 \cdot p_3 \big) - p_{2 \, _\mu}  \big( p_1 \cdot p_3 \big)  \ ,\\
T^{(2)}_\mu (p_1, \, p_2, \,  p_3) & = & - \, T^{(1)}_\mu (p_1, \, p_2, \,  p_3) ~ \big( \slashed{p}_1 - \slashed{p}_2 \big)  \ , \\
T^{(3)}_\mu (p_1, \, p_2, \,  p_3) & = & p^2_3 \, \gamma_\mu - p_{3 \, _\mu} \, \slashed{p}_3  \ , \\
T^{(4)}_\mu (p_1, \, p_2, \,  p_3) & = & T^{(1)}_\mu (p_1, \, p_2, \,  p_3) ~ \sigma_{\alpha\beta} \, p^\alpha_1 \,  p^\beta_2 \ , \\
T^{(5)}_\mu (p_1, \, p_2, \,  p_3) & = & \sigma_{\mu\nu} \, p^\nu_3 \ , \\
T^{(6)}_\mu (p_1, \, p_2, \,  p_3) & = & \gamma_\mu \big( p^2_1 - p^2_2 \big) + \big( p_{1} - p_{2} \big)_\mu \, \slashed{p}_3 \ , \\
T^{(7)}_\mu (p_1, \, p_2, \,  p_3) & = &  - \, \frac{1}{2} \, \big( p^2_1 - p^2_2 \big) \, \big[ \gamma_\mu \,  \big( \slashed{p}_1 - \slashed{p}_2 \big)  - \big( p_{1} - p_{2} \big)_\mu\big] 
               - \big( p_{1} - p_{2} \big)_\mu ~ \sigma_{\alpha\beta} \, p^\alpha_1 \,  p^\beta_2 \ , \\
T^{(8)}_\mu (p_1, \, p_2, \,  p_3) & = &  - \, \gamma_\mu \, \sigma_{\alpha\beta} \, p^\alpha_1 \,  p^\beta_2  \, + \, p_{1 \, _\mu} \slashed{p}_2 \, - \,  p_{2 \, _\mu} \slashed{p}_1  \ ,\label{TensorBasis-Ortho} 
\end{eqnarray}
that is free of kinematical singularities. In Eqs (\ref{TensorBasis-L1}) to (\ref{TensorBasis-Ortho}) we take 
$\sigma_{\mu\nu} = \frac{1}{2} \, [ \gamma_\mu \, , \, \gamma_\nu ]$. 

The longitudinal form factors are determined by the WTI for the vertex, see e.g. \cite{Ball:1980ay} and \cite{Oliveira:2022bar}, and are given by
\begin{eqnarray}
  \lambda_1 (p^2_1, \, p^2_2, \, p^2_3) & = & \frac{1}{2} \bigg( A\big( p^2_1 \big)  + A\big(p^2_2\big) \bigg)  \ ,    \label{EQ:L1} \\
  \lambda_2 (p^2_1, \, p^2_2, \, p^2_3) & = & \frac{1}{2 \, \big( p^2_1 - p^2_2 \big)}   \bigg( A\big( p^2_1 \big)  -  A\big(p^2_2\big) \bigg) \ , \label{EQ:L2} \\
  \lambda_3 (p^2_1, \, p^2_2, \, p^2_3) & = & \frac{1}{ p^2_1 - p^2_2 }   \bigg( B\big(p^2_1\big) - B\big( p^2_2 \big)    \bigg) \ ,   \label{EQ:L3} \\
  \lambda_4 (p^2_1, \, p^2_2, \, p^2_3) & = & 0 \ .  \label{EQ:L4}
\end{eqnarray}
If $A$ and $B$ are smooth functions, then $\lambda_2$ and $\lambda_3$ are regular in the limit of $p^2_1 \rightarrow p^2_2$
and become proportional to the derivatives of $A$ and $B$, respectively, at $p^2_1 = p^2_2$. A straightforward calculation shows that, for zero photon momentum,
\begin{equation}
  \lambda_1 (p^2, \, p^2, 0) = A(p^2) \ , \qquad
  \lambda_2 (p^2, \, p^2, 0) = \frac{1}{2} \,  \frac{d A(p^2)}{d p^2} \qquad\mbox{ and }\qquad
  \lambda_3 (p^2, \, p^2, 0) = \frac{d B(p^2)}{d p^2} \ 
  \label{LongVertex-ZeroMom}
\end{equation}
a result that complies with the Ward identity \cite{Ward:1950xp}
\begin{equation}
 \Gamma^\mu(p, \, p; \, 0) = \frac{\partial S^{-1} (p)}{\partial p_\mu}  \ .
 \label{Wardidentity}
\end{equation}

As discussed, in QED the WTI determine the longitudinal form factors $\lambda_i$.
For non-Abelian theories the WTI identities are replaced by Slavnov-Taylor identities (STI). In QCD, the gluon-fermion vertex verifies a STI that involves Green functions
other than the fermion propagator, namely the ghost dressing function and the quark-ghost scattering kernel.
The solution of this STI for the $\lambda_i$ was worked out in \cite{Aguilar:2010cn}. The expressions found in QCD for the $\lambda_i$ form factors are not as simple as 
in QED.
For example, in QCD the form factor $\lambda_4$ no longer vanishes as happens for the  solution of the vertex WTI in QED. 
The particular case of the solutions of STI in the soft gluon limit, defined by a vanishing gluon momentum, can be found \cite{Oliveira:2018ukh} and should be
compared with the results of Eqs (\ref{LongVertex-ZeroMom}). 
Assuming that the form factors are smooth functions of the momenta and are not singular  for vanishing gluon momentum, 
then the transverse components of the vertex do not contribute to the soft quark-gluon vertex. Indeed, the transverse operators 
that define the vertex basis become null operators in the soft gluon limit.

\section{The full set of Dyson-Schwinger equations \label{Sec:PropEqFF}}

In the previous section a tensor basis to describe the photon-fermion vertex was introduced. The basis allows to write the DSE 
(\ref{DSER-gap}), (\ref{DSER-photon}) and  (\ref{DSER-vertex}) in terms of the form factors $\lambda_i$ and $\tau_i$. 
The  expressions can be found in App. \ref{AppDSEFull}.
In particular,  the scalar and vector components of the fermion gap equation are given in App \ref{Sec:PropEqGap}, the photon equation is reported in
App \ref{Sec:PropEqPhoton} and the decomposition of the photon-fermion vertex in terms of Dirac bilinear forms is described in App. \ref{SecVertex}.

\section{The photon-fermion vertex \label{SecVertexMain}}

The analysis of the DSE for the photon-fermion vertex, see  Eq. (\ref{Eq:DSE-vertex}) for its bare version or (\ref{DSER-vertex}) for the renormalized case,
can be workout with the help of the  basis given in Eqs (\ref{TensorBasis-Long}) and (\ref{TensorBasis-Ortho}). In particular, the l.h.s. of the integral equation
can be written highlighting the tensor decomposition in terms of Dirac bilinears. This decomposition reads
\begin{eqnarray}
& & 
{\Gamma}^\mu (p, \, -p -k; \, k)   ~ = ~ 
     {p}^{\mu } ~  \Big(\, {k}^2 \,  {\tau_1} \,  - \, 2 \, {\lambda_3} \, \Big) - {k}^{\mu }  ~ \Big(\, (pk) \,  {\tau_1} \,  + \, {\lambda_3} \Big)
     \hspace{2.1cm}\text{(Scalar Component)}
     \nonumber \\
     & & 
     \hspace{2cm}
     + \, 
     {\gamma }^{\mu } \Bigg[ \, {\lambda_1} \, + \, k^2 \, {\tau_3}+\,  \Big( k^2  \, + \, \, 2 \,  ( {p} {k}) \Big) \, {\tau_6} \, \Bigg]
      \hspace{3.6cm}\text{(Vector Components)}
     \nonumber \\
     & & \hspace{4cm}
     + \, \slashed{k}  \, \Bigg[ \, {k}^{\mu } \Big( \, {\lambda_2} \, - \,  ({p} {k} ) \,  {\tau_2} \, - \, {\tau_3} \, -\, {\tau_6}\Big) 
                                             \, + \, {p}^{\mu } \Big(2 \, {\lambda_2} \, + \, k^2 \, \, {\tau_2}- \, 2 \, {\tau_6} \Big) \, \Bigg]
     \nonumber \\
     & & \hspace{4cm}
     + \, 2 \, \slashed{p} \, \Bigg[ \, {k}^{\mu } \Big(\, {\lambda_2} \, -  \, ({k} {p}) \, {\tau_2}\Big) \, + \, {p}^{\mu } \Big( 2 \, {\lambda_2} \, + \, k^2 \, {\tau_2} \Big)\,  \Bigg]
     \nonumber \\
     & & \hspace{2cm}
     + \, i ~~ {\tau_8}  ~~ {\gamma }_{\sigma } ~ {\gamma }_5 ~~ ~{\epsilon }^{\sigma  \mu  \alpha\beta} ~~  {p}_\alpha \, {k}_\beta
           \hspace{5.3cm}\text{(Axial Component)}
     \nonumber \\
     & & \hspace{2cm}
     + \,  
     \frac{1}{2} \, \sigma^{\mu\alpha} \, k_\alpha \,  \Bigg[ 2 \, \tau_5 \, + \, \tau_7 \Big( k^2 + 2 \, (kp) \Big) \Bigg]
     \hspace{4.cm}\text{(Tensor Components)}
     \nonumber \\
     & & \hspace{4cm}
     + \,  
          \sigma^{\mu\alpha} \, p_\alpha \,  \Bigg[ \,  \tau_7 \Big( k^2 + 2 \, (kp) \Big) \, \Bigg]
     \nonumber \\
     & & \hspace{4cm}
     \, + \,  \sigma^{\alpha\beta} \, p_\alpha \, k_\beta \Bigg[ \, k^\mu \, \Big( \tau_7 -  (pk) \, \tau_4  \Big) +  p^\mu \Big(  k^2 \, \tau_4  +  2 \, \tau_7  \Big) \Bigg]
     \ ,
     \label{FullVertex-Dirac-main}
\end{eqnarray}
where the form factors are short notations for
\begin{equation}
   \lambda_i = \lambda_i \Big((p+k)^2, \, p^2, \, k^2\Big)
   \qquad\mbox{ and }\qquad
   \tau_i = \tau_i \Big((p+k)^2, \, p^2, \, k^2\Big) \ .
\end{equation}   
As can be read from the above expression, the Dirac scalar bilinear term requires the longitudinal $\lambda_3$, that is fixed by the WTI, and the transverse form factor 
$\tau_1$. Likewise, the Dirac vector term calls for $\tau_2$, $\tau_3$ and $\tau_6$, the Dirac tensor involves $\tau_4$, $\tau_5$, $\tau_7$ and 
the axial vector term is described by a unique form factor $\tau_8$.  
As a cross-check of the decomposition given in Eq. (\ref{FullVertex-Dirac-main}) the contraction $k_\mu \Gamma^\mu$ can be computed and, 
as expected, it demands only the longitudinal form factors $\lambda_i$.

The photon-fermion vertex has been evaluated in perturbation up to one-loop, for a generic linear covariant gauge, in  \cite{Kizilersu:1995iz}. 
In particular, the one-loop perturbative solution for the transverse vertex is known and its tensor structure reproduces, for large photon momentum,
that of the operator associated with $\tau_3$. According to Eq. (\ref{FullVertex-Dirac-main}), for large $k$, the asymptotic transverse vertex reads
\begin{eqnarray}
& & 
{\Gamma}^\mu_T (p, \, -p -k; \, k)   ~ = ~ 
     {p}^{\mu } ~  \, {k}^2 \,  {\tau_1}   - {k}^{\mu }  ~ \, (pk) \,  {\tau_1} 
     ~ + ~ 
     {\gamma }^{\mu } \bigg( \, k^2 \, {\tau_3} +   \,  k^2   \, {\tau_6} \, \bigg)
     \nonumber \\
     & & \qquad
     + \, \slashed{k}  \, \Bigg[ \,  {p}^{\mu } \Big(  \, k^2 \, \, {\tau_2}- \, 2 \, {\tau_6} \Big) \,  
                                                - \,  {k}^{\mu } \Big( \,   ({p} {k} ) \,  {\tau_2} \, + \, {\tau_3} \, + \, {\tau_6}\Big)\Bigg]
     ~ + ~ 2 \, \slashed{p} \, \Bigg[ \,  \, {p}^{\mu }  \, k^2 \, {\tau_2} \, - {k}^{\mu } \,  ({k} {p}) \, {\tau_2} \,  \Bigg]
     \nonumber \\
     & & \qquad
     + \, i ~~ {\tau_8}  ~~ {\gamma }_{\sigma } ~ {\gamma }_5 ~~ ~{\epsilon }^{\sigma  \mu  \alpha\beta} ~~  {p}_\alpha \, {k}_\beta
     \nonumber \\
     & & \qquad
     + \,  
     \frac{1}{2} \, \sigma^{\mu\alpha} \, k_\alpha \,  \Bigg[ 2 \, \tau_5 \, + \, \tau_7 \,  k^2  \, \Bigg]
     ~+ ~  
          \sigma^{\mu\alpha} \, p_\alpha \,  \Bigg[ \,  \tau_7 \, k^2  \, \Bigg]
     ~ + ~  \sigma^{\alpha\beta} \, p_\alpha \, k_\beta \Bigg[ \, k^\mu \, \Big( \tau_7 -  (pk) \, \tau_4  \Big) +  p^\mu \Big(  k^2 \, \tau_4  +  2 \, \tau_7  \Big) \Bigg]
     \nonumber \\
     & & \qquad
      = \frac{\alpha \, \xi}{8 \, \pi} \left( \frac{k^\mu \slashed{k}}{k^2}  -  \gamma^\mu \right) \, \ln \frac{k^2}{p^2} \ ,
     \label{FullVertex-Dirac-UV-transverse}
\end{eqnarray}
where, in the last line, the asymptotic one-loop perturbative result is quoted. The recovering of the perturbative expression requires that,
for large $k$,
\begin{equation}
\tau_3 \approx - \, \frac{\alpha \, \xi}{8 \, \pi \, k^2}  \, \ln \frac{k^2}{p^2} \ , \qquad
\tau_{1, \, 6} < \frac{1}{k^2} \ , \qquad
\tau_{5} < \frac{1}{k} \ , \qquad
\tau_{2, \, 4, \, 7} < \frac{1}{k^3} \ , \qquad\mbox{ and }\qquad
\tau_{8} < \frac{1}{k} \ .
\label{PertVertex-oneloop}
\end{equation}
In what concerns the form factors $\tau_{i \ne 3}$, their exact asymptotic scaling law is not determined exactly. 
The above relations define only upper limits on their asymptotic scaling with the photon momentum.

Profiting from the decomposition of the vertex in Dirac bilinears,  the contraction of the vertex equation with the incoming fermion momentum $p^\mu$ 
results in another Lorentz scalar that can be used to access $\tau_1$. Indeed, this contraction gives
\begin{eqnarray}
& & 
p_\mu \,{\Gamma}^\mu (p, \, -p -k; \, k)   ~ = ~ 
     \tau_1 ~ \Big( \, p^2 \, k^2 - (pk)^2 \,\Big) \, - \, \lambda_3 ~ \Big( \, 2 \, p^2 + (pk) \, \Big) 
     \nonumber \\
     & & 
     \qquad\qquad
     + \, 
     \slashed{k} \Bigg[ \, 
              \lambda_2 \bigg( (kp) + 2 \, p^2 \bigg)
              ~ + ~ \tau_2 \bigg( k^2p^2 - (kp)^2 \bigg)
              ~ - ~ \tau_3 \, (kp) 
              ~ - ~ \tau_6 \bigg( (kp) + 2 \, p^2 \bigg)
      \, \Bigg]
     \nonumber \\
     & & \qquad\qquad\qquad
     + \,  \slashed{p} \, \Bigg[ \, 
          \lambda_1
          ~ + ~ 2 \, \lambda_2 \bigg( (kp) + 2 \, p^2 \bigg)
          ~ + ~ 2 \, \tau_2 \bigg( k^2 p^2 - (kp)^2 \bigg)
          ~ + ~ \tau_3 \, k^2 
          ~ + ~ \tau_6 \bigg( k^2 + 2 \, (kp) \bigg) 
     \,  \Bigg]
     \nonumber \\
     & & \qquad\qquad
     + \,  
     \sigma^{\alpha\beta} \, p_\alpha \, k_\beta \Bigg[ \, 
      \tau_4 \bigg( p^2 k^2 - (kp)^2 \bigg)
      ~ + ~ \tau_5 
      ~ + ~ \tau_7 \bigg( \frac{1}{2} k^2 + 2 \, (kp) +  2 \, p^2 \bigg)       
     \, \Bigg] \ .
     \label{FullVertex-Dirac-Contraction}
\end{eqnarray}
Recalling that the longitudinal form factors are determined by the Ward-Takahashi identity for the vertex, then  $\tau_1$ can be evaluated
solving the equation
\begin{equation}
  \tau_1  \Big((p+k)^2, \, p^2, \, k^2\Big)=   ~ \frac{1}{ p^2 \, k^2 - (pk)^2 } \Bigg( 
\frac{1}{4} \, \text{Tr} \bigg( p_\mu \,{\Gamma}^\mu (p, \, -p -k; \, k)   \bigg) +  \lambda_3 ~ \Big( \, 2 \, p^2 + (pk) \, \Big) 
\Bigg) \ .
\label{tau1-fromvertex}
\end{equation}   
This is an exact relation and is an example of the type of relations that will be considered to access the transverse form factors.
These relations explore the tensorial properties of $\Gamma^\mu$. 
Another example of such type for the photon-fermion vertex allows the computation of $\tau_8$ is 
\begin{equation}
~~ {\tau_8}  \Big((p+k)^2, \, p^2, \, k^2\Big) ~~{\epsilon }^{\zeta  \mu  \alpha\beta} ~~  {p}_\alpha \, {k}_\beta ~ = ~
i ~~ \frac{1}{4} \text{Tr} \Bigg( {\Gamma}^\mu (p, \, -p -k; \, k)   \, \gamma^\zeta \, \gamma_5 \Bigg)  \ .
\label{tau8-fromvertex}
\end{equation}
This form factor  vanishes for a tree level type of vertex.
Indeed, for a tree level type of vertex where $\Gamma^\mu = \gamma^\mu$, $\lambda_2 = \lambda_3 = 0$ and the above equations imply that $\tau_1 = \tau_8 = 0$.

\subsubsection{Special kinematics}

The vertex decomposition in terms of Dirac bilinears given in Eq. (\ref{FullVertex-Dirac-main}) is valid for any kinematical configuration. 
However, for particular choices of the incoming quark momentum $p$ and of the photon incoming momentum $k$, assuming that the form factors are not singular, 
the decomposition simplifies.
For example, for a vanishing  incoming fermion momentum $p = 0$ it comes that
\begin{eqnarray}
{\Gamma}^\mu (0, \, -k; \, k)   & = &
     - {k}^{\mu }  ~  {\lambda_3} 
    ~  + ~ 
     {\gamma }^{\mu } \Bigg[ \, {\lambda_1} \, + \, k^2 \, \Big( \tau_3   \, + \,  {\tau_6} \Big) \, \Bigg]
     ~ + ~ \slashed{k}  ~ {k}^{\mu } \Bigg[ \, {\lambda_2}  \, - \, {\tau_3} \, -\, {\tau_6}\Bigg]
     \nonumber \\
     & & \qquad\qquad
     + \,  
     \frac{1}{2} \, \sigma^{\mu\alpha} \, k_\alpha \,  \Bigg[ 2 \, \tau_5 \, + \, k^2  \, \tau_7  \Bigg] \ ,
     \label{FullVertex-Dirac-pzero}
\end{eqnarray}
and it is enough to know the combinations of transverse form factors $\tau_3 + \tau_6$ and $2 \, \tau_5 \, + \, k^2  \, \tau_7$ to describe $\Gamma^\mu$.
Recall that the longitudinal form factors are the solutions of the vertex WTI.

Another kinematical configuration where Eq. (\ref{FullVertex-Dirac-main}) simplifies considerably is the case of a vanishing photon momentum (soft photon limit),
where $\Gamma^\mu$ is described only by the longitudinal form factors $\lambda_i$. Indeed, in this kinematics, and as long as the $\tau_i$ do not have poles
at zero momentum, the expression for the vertex become
\begin{eqnarray}
   {\Gamma}^\mu (p, \, -p ; \, 0)   & = &
   \, 2 \, \lambda_3 (p^2, \, p^2, 0)  ~ p^\mu ~ + ~    \lambda_1 (p^2, \, p^2, 0)  ~  {\gamma }^{\mu } ~ +  ~  4 \, \lambda_2 (p^2, \, p^2, 0) 
  ~ \slashed{p} \, p^\mu  
\nonumber \\
&  = & - \, 2 \, B^\prime (p^2)  ~ p^\mu ~ + ~    A(p^2)  ~  {\gamma }^{\mu } ~ +  ~  2 \, A^\prime (p^2)  ~ \slashed{p} \, p^\mu \ .
\nonumber \\
& = &   
    \gamma^\mu  ~ + ~ i \, g^2 \, \int \frac{d^4q}{(2 \, \pi)^4} ~D_{\zeta\zeta^\prime}(q)  ~ \gamma^\zeta ~ S(p-q)
   \nonumber \\
& & 
\qquad\qquad
   \Bigg\{ ~ 
                                 \, {\Gamma}^\mu ( p - q, \, -p + q; \, 0) \,
                                S(p-q) \, {\Gamma}^{\zeta^\prime} (p -q, \, -p; q) 
~ + ~
                                 \, {\Gamma}^{\zeta^\prime\mu} (p - q, \, -p; \, q , \, 0) 
                                \Bigg\}  \ .
                                \label{Eq:DSE-vertex-soft}
\end{eqnarray}
In the above expression, the result of Eq. (\ref{LongVertex-ZeroMom}) was used to write the longitudinal form factors in terms of the fermion propagator functions.
The (bare) vertex DSE is algo given explicitly to identify the various contributions to $\Gamma^\mu$.
The two vertices ${\Gamma}^\mu (p, \, -p ; \, 0)$ and $ {\Gamma}^\mu ( p - q, \, -p + q; \, 0)$, that appears in the r.h.s. of the DSE equation,
require only the longitudinal form factors, see the first line of Eq. (\ref{Eq:DSE-vertex-soft}),  and are determined by the vertex WTI. 
On the other hand, the longitudinal component of the two-photon-two-fermion vertex for the kinematics appears also in the above equation.
The solution of the corresponding WTI writes $\Gamma^{\mu\nu}$ in terms of the fermion propagator functions $A$ and $B$ and their derivatives, and
can be found in \cite{Oliveira:2022bar}.
For a general kinematics, the two-photon-two-fermion vertex $\Gamma^{\mu\nu}$ has a complex tensor structure that includes both longitudinal and transverse, relative
to the photons momenta, tensor operators.
For completeness, we provide the expression for the two-photon-two-fermion derived in \cite{Oliveira:2022bar} that solves the associated WTI
when one of the photon momentum vanishes
\begin{eqnarray}
\Gamma^{\zeta\mu} (p-q, \, -p; \, q, \, 0) & = &
  2 ~ g^{\zeta\mu} ~ B^\prime(p^2) ~ + ~  2 ~ \frac{q^\zeta p^\mu - q^\zeta q^\mu}{q^2}  ~ \Bigg[ B^\prime (p^2) - B^\prime\Big((p-q)^2\Big) \Bigg]
  \nonumber \\
  & & \quad
  ~ + ~ \frac{q^\zeta \gamma^\mu}{q^2} ~ \Bigg[  A\Big((p-q)^2\Big) - A (p^2)   \Bigg]
  \nonumber \\
  & & \quad\quad
  ~ - ~ 2 ~ \slashed{q} ~ \frac{q^\zeta p^\mu - q^\zeta q^\mu}{q^2} ~  A^\prime\Big((p-q)^2\Big) 
  \nonumber \\
  & & \quad\quad\quad
    ~ - ~ 2 ~ \slashed{p} \Bigg\{ ~ g^{\zeta\mu} ~A^\prime(p^2) ~ + ~ \frac{q^\zeta p^\mu - q^\zeta q^\mu}{q^2}  ~ \Bigg[ A^\prime (p^2) - A^\prime\Big((p-q)^2\Big) \Bigg] ~ \Bigg\} \ ,
    \label{Eq:Gamma-mu-nu-q0}
\end{eqnarray}
where the expressions derived previously for $\lambda_i$ were used. 
If the inverse of the fermion wave function has a weak dependence on the momentum,
an approximation that is often used by setting $A (p^2) = 1$, then the last terms in $\Gamma^{\zeta\mu}$ can be ignored and
\begin{eqnarray}
\Gamma^{\zeta\mu} (p-q, \, -p; \, q, \, 0) & \approx &
  2 ~ g^{\zeta\mu} ~ B^\prime(p^2) ~ + ~  2 ~ \frac{q^\zeta p^\mu - q^\zeta q^\mu}{q^2}  ~ \Bigg[ B^\prime (p^2) - B^\prime\Big((p-q)^2\Big) \Bigg] \ ,
    \label{Eq:Gamma-mu-nu-q0-approx}
\end{eqnarray}
and vanishes at high momenta, where an essentially constant $B(p^2)$ is expected to match the perturbative result.

Another kinematical configuration is when the incoming fermion momentum and the photon momentum  are orthogonal, i.e. when $(pk) = 0$. 
In this case, the vertex become
\begin{eqnarray}
& & 
{\Gamma}^\mu (p, \, -p -k; \, k)   ~ = ~ 
     {p}^{\mu } ~  \Big(\, {k}^2 \,  {\tau_1} \,  - \, 2 \, {\lambda_3} \, \Big) - {k}^{\mu }  ~  {\lambda_3} 
     \nonumber \\
     & & 
     \qquad\qquad
     + \, 
     {\gamma }^{\mu } \Bigg[ \, {\lambda_1} \, + \, k^2 \, \Big( {\tau_3} \, + \,   {\tau_6} \Big) \, \Bigg]
     ~ + ~ \slashed{k}  \, \Bigg[ \, {k}^{\mu } \Big( \, {\lambda_2} \, - \, {\tau_3} \, -\, {\tau_6}\Big) 
                                             \, + \, {p}^{\mu } \Big(2 \, {\lambda_2} \, + \, k^2 \, \, {\tau_2}- \, 2 \, {\tau_6} \Big) \, \Bigg]
     \nonumber \\
     & & \qquad\qquad\qquad
     + \, 2 \, \slashed{p} \, \Bigg[   {p}^{\mu } \Big( 2 \, {\lambda_2} \, + \, k^2 \, {\tau_2} \Big)\, \, - \,  {k}^{\mu } \, {\lambda_2}  \Bigg]
     \nonumber \\
     & & \qquad\qquad
     + \, i ~~ {\tau_8}  ~~ {\gamma }_{\sigma } ~ {\gamma }_5 ~~ ~{\epsilon }^{\sigma  \mu  \alpha\beta} ~~  {p}_\alpha \, {k}_\beta
     \nonumber \\
     & & \qquad\qquad
     + \,  
     \frac{1}{2} \, \sigma^{\mu\alpha} \, k_\alpha \,  \Bigg[ 2 \, \tau_5 \, + \, \tau_7 \, k^2  \Bigg]
      ~ + ~  
          \sigma^{\mu\alpha} \, p_\alpha \,  \Bigg[ \,  \tau_7 \, k^2  \, \Bigg]
     ~  + ~ \sigma^{\alpha\beta} \, p_\alpha \, k_\beta \Bigg[ \, k^\mu \, \tau_7  +  p^\mu \Big(  k^2 \, \tau_4  +  2 \, \tau_7  \Big) \Bigg]
     \label{FullVertex-Dirac-orthogonal}
\end{eqnarray}
and, again, only particular combinations of the transverse form factors have to be considered.

\subsubsection{Comment on vertex models}

From a practical point of view, the use of the DSE to solve QED or QCD rely, many times, on modelling the fermion-boson vertex. The call for vertex models
necessarily lead to a simplification of Eq. (\ref{FullVertex-Dirac-main}) tensor structure.  
A subset of the available vertex models that can be found in the literature is
\cite{Curtis:1990zs,Curtis:1992jg,Roberts:1994dr,Maris:1999nt,Kizilersu:2009kg,Kizilersu:2014ela,Lessa:2022wqc}; see also the references therein. 
For example, the Curtis-Pennington vertex \cite{Curtis:1990zs,Curtis:1992jg} considers
\begin{eqnarray}
{\Gamma}^\mu (p, \, -p -k; \, k)   & =  & 
       \Big(\,  - \, 2 \, {p}^{\mu } ~  - ~  {k}^{\mu } \, \Big)  ~ {\lambda_3} 
     ~ + ~ 
     {\gamma }^{\mu } \Bigg( \, {\lambda_1} \, + \,  \Big( k^2  \, + \,  2 \,  ({k} {p}) \, \Big) \, {\tau_6}  \, \Bigg) 
     \nonumber \\
     & & \qquad
     + \, \slashed{k}  \, \Bigg( \, {k}^{\mu }  \, + \, 2 \, {p}^{\mu }  \, \Bigg) 
                                             \Big( \, {\lambda_2}  \, -\, {\tau_6}\Big) 
    ~  + ~ 2 \, \slashed{p} \, \Bigg( \, {k}^{\mu }  \, + \, 2 \, {p}^{\mu }  \,  \Bigg)
     \, {\lambda_2} \ .
     \label{FullVertex-Dirac-CP}
\end{eqnarray}
This vertex model for QED is compatible with multiplicative renormalizability, for massless fermions and in the quenched approximation,
and considers a single transverse form factor $\tau_6$. This form factor  contributes to  $B(0)$, see Eq. (\ref{DSE-Fermion-scalar}),  and, therefore,
can be made compatible with the chiral symmetry breaking mechanism for a large enough coupling constant.

As a final comment, we note that the decomposition discussed for the vertex, or any of its simplifications, explore the tensor structure of $\Gamma^\mu$. 
In general, they are valid both for the photon-fermion vertex and also for the quark-gluon vertex, module its color structure and after including the contribution 
of $\lambda_4$. 
The same rationale applies to the decomposition of the fermion gap equation in terms of the vertex form factors,  see Eqs (\ref{DSE-Fermion-scalar}) and 
(\ref{DSE-Fermion-vector}). Indeed, the QCD fermion gap equation, after taking into account the color factors and $\lambda_4$, that appears only in the scalar equation, 
is the same up to a proper replacement of the coupling constant and the renormalization constants. 
The main difference between QCD and QED is the bosonic DSE that, in QCD, require gluon one-particle irreducible Green functions
with three and four gluon legs together with ghost contributions.

\section{The on-shell photon-fermion vertex\label{Sec:OnShell}}

The particular case that corresponds to the evaluation of the vertex diagram with fermions that are on-shell allows the use of the Dirac
equation to simplify the writing of the vertex and build an effective on-shell vertex. Let us define the on-shell photon-fermion vertex $\widetilde{\Gamma}$ by
\begin{equation}
    \Big[ \bar u (p^\prime) ~  \widetilde{\Gamma}^\mu (p, \, -p^\prime = - p - k; \, k) ~  u(p) \Big]  = 
    \left. 
   \Big[ \bar u (p^\prime) ~  \Gamma^\mu (p, \, -p^\prime = - p - k; \, k) ~  u(p) \Big]
    \right|_{p^2=p^{\prime \, 2} = m^2} \ .
\end{equation}   
Assuming that the free Dirac equation can be applied to simplify the r.h.s of this equation, then, after some algebra, one arrives at
\begin{eqnarray}
\widetilde{\Gamma}^\mu (p, \, -p^\prime = - p - k; \, k)  & = &
\Big(p^\prime + p\Big)^\mu ~
    \Bigg\{
    - \, \lambda_3 + \Big( m^2 - (p^\prime p) \Big) \, \tau_1 + \tau_5  + 2 \, m \Big[ \lambda_2 + \Big( m^2 - (p^\prime p) \Big) \tau_2 \Big]
    \Bigg\}
    \nonumber \\
    & & \qquad
    + ~ \gamma^\mu ~ \Bigg\{
        \lambda_1 + 2 \, \Big( m^2 - (p^\prime p) \Big) \, \tau_3 - 2 \, m \, \tau_5
    \Bigg\}
    \nonumber \\
    & & \qquad\qquad
    + \, i ~~ {\tau_8}  ~~ {\gamma }_{\sigma } ~ {\gamma }_5 ~~ ~{\epsilon }^{\sigma  \mu  \alpha\beta} ~~  {p}_\alpha \, p^\prime_\beta
    \nonumber \\
    & = & 
    \gamma^\mu ~
    \Bigg\{
    \lambda_1 + 2 \, \Big( m^2 - (p^\prime p) \Big) \, \tau_3
             + 2 \, m \, \Bigg[
                              2 \, m \, \lambda_2  \, - \,  \lambda_3 + \Big( m^2 - (p^\prime p) \Big) \,  \Big( \tau_1 + 2 \, m \, \tau_2 \Big)
                              \Bigg]
    \Bigg\}
    \nonumber \\
    & & \qquad
    + ~
     \sigma^{\mu\alpha} k_\alpha ~ \Bigg\{
           - \, \lambda_3 + \Big( m^2 - (p^\prime p) \Big) \, \tau_1 + \tau_5  + 2 \, m \Big[ \lambda_2 + \Big( m^2 - (p^\prime p) \Big) \tau_2 \Big] 
     \Bigg\}
     \nonumber \\
     & & \qquad\qquad
         + \, i ~~ {\tau_8}  ~~ {\gamma }_{\sigma } ~ {\gamma }_5 ~~ ~{\epsilon }^{\sigma  \mu  \alpha\beta} ~~  {p}_\alpha \, p^\prime_\beta \ ,
     \label{OnShellVertex}
\end{eqnarray}
where the last expression was obtained from the first with the help of the Gordon identity
\begin{equation}
   2 \, m \, \bar u(p) \gamma^\mu u(p) = 
    \bar u(p) \bigg( \left( p^\prime + p \right)^\mu - \sigma^{\mu\alpha} \left( p^\prime - p \right)_\alpha  \bigg)u(p) \ .
\end{equation} 
An interesting limit of this vertex is the chiral on-shell photon-fermion vertex, that corresponds to take the limit $m \rightarrow 0$ together with $\lambda_3 \rightarrow 0$,
as the solutions of the vertex WTI imply. Assuming a smooth behaviour of the form factors it follows that
\begin{eqnarray}
\widetilde{\Gamma}^\mu (p, \, -p^\prime = - p - k; \, k)  & = &
    \gamma^\mu ~
    \bigg(
    \lambda_1 - 2 \,  (p^\prime p)  \, \tau_3
    \bigg)
~ + ~
     \sigma^{\mu\alpha} k_\alpha ~ \bigg(
      \tau_5  - (p^\prime p) \, \tau_1 
     \bigg)
~  + ~ i ~ {\tau_8}  ~~ {\gamma }_{\sigma } ~ {\gamma }_5 ~~ ~{\epsilon }^{\sigma  \mu  \alpha\beta} ~~  {p}_\alpha \, p^\prime_\beta  ,
  \label{OnShellVertexChiral}
\end{eqnarray}
i.e. the on-shell vertex in the chiral gets contributions only from $\lambda_1$, $\tau_1$, $\tau_3$, $\tau_5$ and $\tau_8$. For the vertex models that do not consider these 
transverse form factors,  the on-shell vertex in the chiral limit is reduced to its tree level tensor structure.

The anomalous magnetic and electric fermion form factors can be read from Eq. (\ref{OnShellVertex}) or, for their chiral limit, 
from Eq. (\ref{OnShellVertexChiral}).
Note that the vertex Ward-Takahashi identity for the photon-fermion vertex implies
\begin{eqnarray}
k_\mu ~ \Big[ \bar u (p^\prime) ~  \widetilde{\Gamma}^\mu (p, \, -p^\prime = - p - k; \, k) ~  u(p) \Big] & = &
k_\mu ~ \Big[ \bar u (p^\prime) ~  {\Gamma}^\mu (p, \, -p^\prime = - p - k; \, k) ~  u(p) \Big] 
\nonumber \\
& = &  \Bigg[ \bar u (p^\prime) ~  \bigg( S^{-1}(-p^\prime) -  S^{-1}(p) \bigg) ~  u(p) \Bigg] 
\nonumber \\
& = &  \Bigg[ \bar u (p^\prime) ~  \bigg( B(p^2) - B(p^{\prime \, 2})  - m  \bigg( A(p^{\prime \, 2}) + A(p^2) \bigg)  \bigg) ~  u(p) \Bigg] 
\end{eqnarray}
that vanishes either in the chiral limit or when $A$ and $B$ are momentum independent functions, as occurs for the tree level solution of QED.

The results derived in this section are quite general and are  valid also for QCD, after the correction of the color structure and after taking into
consideration the contribution associated with $\lambda_4$. 
If for QED the on-shell condition seems to be a reasonable approximation, in QCD the on-shell condition rises conceptual difficulties due
quark confinement.
There are also important differences between the longitudinal form factors for QED and QCD.
The Ward-Takahashi identity for the vertex in QED is replaced, in QCD, by a Slavnov-Taylor identity (STI),
that has contributions from the quark-ghost scattering kernel and ghost dressing function. The solution of the vertex STI for the $\lambda_i$ 
has a higher complexity when compared to the Abelian vertex.
However, in lowest order in the coupling constant, the perturbative solution of the Slavnov-Taylor identity for the vertex reproduces the results 
for the longitudinal vertex of the Abelian theory and, in this sense, it is tempting to use the Ball-Chiu vertex in QCD, despite its  known limitations.
A discussion on the longitudinal form factors in QCD can be found in e.g.
\cite{Oliveira:2018fkj,Alkofer:2000wg,Fischer:2006ub,RichardWilliams2007,Aguilar:2010cn,Aguilar:2016lbe,Binosi:2016wcx,Aguilar:2018epe,Aguilar:2018csq,Oliveira:2018ukh,Oliveira:2020yac} and references therein.
See also \cite{Mena:2023mqj} for a recent discussion of the QCD corrections to the on-shell photon-quark vertex.

\section{Transverse Form Factors from the Photon-Fermion vertex DSE \label{SecTransverseFF}}
 
 In this section a computational framework for the evaluation of the transverse form factors $\tau_i$ from the vertex Dyson-Schwinger equation is built. 
 Such procedure was already worked out for $\tau_1$, see Eq. (\ref{tau1-fromvertex}), and for $\tau_8$ in Eq. (\ref{tau8-fromvertex}). 
 It remains to derive exact expressions for the remaining six form factors. The procedure rely on the photon-fermion vertex DSE, 
 i.e on Eq. (\ref{DSER-vertex}), and the tensor decomposition of the vertex as given in Eqs (\ref{FullVertex-Dirac-main}) and  (\ref{FullVertex-Dirac}). 
 The equations to be written below should be solved self-consistently for the transverse form factors and are generalizations of those derived for $\tau_1$ and $\tau_8$.
 We start by reviewing the procedure for these two form factors.

\subsection{The form factor $\tau_1$ \label{SubSecTau1}}

The transverse form factor $\tau_1$ can be computed using Eq. (\ref{tau1-fromvertex}) and it is given by
the solution of equation
\begin{equation}
  \tau_1  \Big((p+k)^2, \, p^2, \, k^2\Big)=   ~ \frac{1}{ p^2 \, k^2 - (pk)^2 } \Bigg\{
\frac{1}{4} \, \text{Tr} \bigg( p_\mu \,{\Gamma}^\mu (p, \, -p -k; \, k)   \bigg) ~ + ~ \frac{ 2 \, p^2 + (pk) }{ k^2 + 2 \, (pk) } ~ \Bigg( B((p+k)^2) - B( p^2) \Bigg) 
\Bigg\} \ .
\label{tau1-fromvertex-Mink}
\end{equation}   
It is instructive to compare this expression for $\tau_1$ with the result obtained in perturbation theory at one-loop level for QED \cite{Kizilersu:1995iz}.
When reading the expressions from this work, 
the different momentum convention\footnote{For example,
in \cite{Kizilersu:1995iz} the quantity $\Delta^2$ corresponds, in our conventions, to $(pk)^2 - p^2 k^2$ that later will be called $- \, \Delta$.} should
be taken into account.
In both cases, i.e. in the exact  expression given in (\ref{tau1-fromvertex-Mink}) and in the outcome of one-loop perturbation theory,
the overall factor $1/( p^2 \, k^2 - (pk)^2)$ is seen and defines a scaling law for $\tau_1$. 
The comparison also allows for an estimation of the trace term written in the above equation.

Oftentimes, in a practical situation, the vertex is modelled and a given tensor structure is considered explicitly. 
The overall factor mention previously is not always considered in the models for $\tau_1$.
See, for example, \cite{Albino:2021rvj,El-Bennich:2022obe,Lessa:2022wqc} and references therein for models that consider the overall kinematical factor. 
An example where the kinematical factor it is not used can be found e.g. in \cite{Bashir:2011dp}. Other vertex models do not even consider
$\tau_1$ at all as, for example, in \cite{Curtis:1990zs,Curtis:1992jg,Kizilersu:2009kg,Aguilar:2012rz,Oliveira:2020yac}.

Note also that, at lowest order in perturbation theory, i.e. for vertex $\Gamma^\mu = \gamma^\mu$ and a $B$ that is momentum independent, this form factor vanish. 
This result agrees with the chiral limit of the one-loop perturbative vertex calculation, see \cite{Kizilersu:1995iz}, where $\tau_1$ is proportional to the 
fermion mass. Also it implies that, in perturbation theory, this form factor is, at large momenta, necessarily small for light fermions.

\subsection{The form factor $\tau_8$}

For completeness we review the computation of $\tau_8$ that is associated with the Dirac axial vector of the decomposition of $\Gamma^\mu$, see 
Eq.  (\ref{FullVertex-Dirac-main}) or (\ref{FullVertex-Dirac}). It follows that  
 \begin{equation}
      {\tau_8}  ~~ {\epsilon }^{\zeta  \mu  \alpha\beta} ~~  {p}_\alpha \, {k}_\beta =  ~
       \, \frac{i}{4} \mbox{Tr} \Big( \gamma^\zeta \, \gamma_5 \, \Gamma^\mu \Big)
       \label{GetTau8}
 \end{equation}
 and after contraction with the Levi-Civita tensor and momenta it is possible to arrives at a scalar equation for $\tau_8$.
 Further discussion on the computation of $\tau_8$ can be found in Sec. \ref{Sec:PertFFactors}.

\subsection{The form factors $\tau_2$, $\tau_3$ and $\tau_6$ \label{SecGetTau236}}

The transverse form factors $\tau_2$, $\tau_3$ and $\tau_6$
can be computed from Eq. (\ref{FullVertex-Dirac-main})  combined with the vectorial part of Eq. (\ref{FullVertex-Dirac-Contraction}).
To disentangle the various form factors,  let us define
\begin{eqnarray}
\!\!
\Lambda_1 & = &  \frac{1}{4} \, \mbox{Tr} \left( \gamma_\mu \Gamma^\mu \right)  
                   - 4  \,  \lambda_ 1 -  \Big( k^2  + 4 (pk)  + 4 p^2 \Big) \, \lambda_2  
                   ~ ~  = \,
                   2 \, \Big( p^2  k^2  -   (pk)^2 \Big) \, \tau_2   + 3 \,  k^2 \, \tau_3 + 3 \, \Big( k^2 + 2 \, (pk) \Big) \, \tau_6  \ ,
   \\      
\!\!
\Lambda_2 & = &    \frac{1}{4} \, \mbox{Tr} \left( \slashed{k} \,  \gamma_\mu \Gamma^\mu \right)
                  -  (pk) \, \lambda_1 - \Big(  2 \, p^2 + (p k) \Big) \Big( k^2 + 2  (p k) \Big) \, \lambda_2  
                  ~ ~  =   \Big( p^2 k^2 - (pk)^2 \Big) \, \bigg[
                  \Big( k^2 + 2 (pk) \Big)   \,  \tau_2 \, -  \, 2 \,   \tau_6 \bigg] \ ,
 \\      
\!\!
\Lambda_3 & = &    \frac{1}{4} \, \mbox{Tr} \left( \slashed{p} ~ \gamma_\mu \Gamma^\mu \right)                  
                   -  p^2 \, \lambda_1 -  \Big( 2 \, p^2 + (pk) \Big)^2 \, \lambda_2
                  ~~ = \, \Big( p^2 k^2 - (pk)^2 \Big) \, \bigg[
                   \Big( 2 \, p^2 + (pk) \Big)  \, \tau_2 \,+ \,  \tau_3 \, + \,  \tau_6 \bigg] \ .
\end{eqnarray}
Solving these identities in terms of the transverse form factor, it follows that 
\begin{eqnarray}
\tau_2 & = &
    - \,\frac{1}{4 \, \Delta} \Bigg( ~  \Lambda_1 ~ + ~ 3 \, \frac{(pk)}{\Delta} \, \Lambda_2 ~ - ~ 3 \, \frac{k^2 }{\Delta} \, \Lambda_3  ~ \Bigg)
    \ ,
     \label{FFtau2:Exact}
\\
\tau_3 & = & 
    \frac{1}{8 \, \Delta^2}  \Bigg( ~
         \Delta \, 
           \bigg(  k^2  \,   + \, 4 \,  p^2  \, +  \, 4 \,  (pk)    \bigg) \, \Lambda_1 
~ +   ~   \bigg(  4 \Delta \, + \, 3 \, (pk) \Big( k^2 + 4 \, p^2 + 4 \, (pk)^2\Big)  \bigg) \, \Lambda_2
       \nonumber \\
       & & \qquad\qquad\qquad
~ -  ~  \bigg(   4 \, \Delta \, + \, 3 \, k^4  + 12 \, (pk) \Big( k^2 + (pk) \Big)  \bigg) \, \Lambda_3 ~ \Bigg) 
\,  ,
     \label{FFtau3:Exact}
\\
\tau_6 & = &
\frac{1}{8 \, \Delta^2} ~ \Bigg( ~
  - \, \Delta  \,  \bigg(  k^2 + 2 \, (pk) \bigg)\,  \Lambda_1
~ -   ~    \bigg(  4 \, \Delta + 3 \, (pk) \Big( k^2 +  2 \, (pk) \Big) \bigg) \, \Lambda_2
~ +  ~ 3 \, k^2 \,  \bigg( k^2 + 2 \,  (pk) \bigg) \, \Lambda_3 ~
\Bigg) \ .
     \label{FFtau6:Exact}
\end{eqnarray}
As for the form factors $\tau_1$ and $\tau_8$, the above relations are equations that have to be solved self-consistently with respect to
 $\tau_2$, $\tau_3$ and $\tau_6$ as $\Gamma^\mu$, or its contractions, are  functions of all the vertex form factors. Once more, we recall the reader that
 the longitudinal form factors $\lambda_i$ are determined by the WTI and are functions of the fermion propagator form factors $A$ and $B$.

As for $\tau_1$ and $\tau_8$, all the form factors $\tau_2$, $\tau_3$ and $\tau_6$ share the common factor multiplicative kinematical factor
\begin{equation}
   \frac{1}{\Delta} = \frac{1}{ p^2 k^2 - (pk)^2 } 
   \label{FFDelta}
\end{equation}  
that can come as a higher power of $\Delta$.
This factor is also present in the one-loop perturbative solution for these transverse form factors.
This common factor is not always taken into account in the models for the photon-fermion vertex. 
For a tree level vertex type the $\Lambda_i$  vanish and, therefore, all the form factors $\tau_2$, $\tau_3$ and $\tau_6$ vanish. 
From the above expression one can derive naive asymptotic dependences on the photon momentum that should be read with some care as the
$\Lambda_i$ are also functions of the form factors themselves.
A naive looking at the above expressions suggests that for large photon momentum the
expressions are dominated by $\tau_2$ and are proportional to $\Lambda_1$.

\subsection{The form factors $\tau_4$, $\tau_5$ and $\tau_7$}

Proceeding as in the previous section but looking at the tensor components of Eq. (\ref{FullVertex-Dirac-main})  one can access
the  transverse form factors $\tau_4$, $\tau_5$ and $\tau_7$ from 
\begin{eqnarray}
 \Omega_1 & = & 
                \frac{1}{4} \, \mbox{Tr} \Big( \sigma^{\alpha\beta} p_\alpha k_\beta ~ \big( p_\mu \Gamma^\mu \big) \Big) 
                ~ = \, -  \, 
                \big(p^2 k^2 - (pk)^2\big) \Bigg( 
                 \bigg( p^2 k^2 - (pk)^2\bigg) \, \tau_4    + \tau_5 \, 
                 +  \frac{1}{2} \,   \bigg( k^2 + 4 \, p^2 + 4 \, (pk) \bigg)  \, \tau_7 \Bigg)  ,
                \label{FFtau4-Omega1}
 \\
 \Omega_2 & = & 
             \frac{1}{4} \, \mbox{Tr} \Big( \sigma_{\alpha\mu} \, k^\alpha ~ \Gamma^\mu  \Big) 
                ~ = \,
                k^2 \, \bigg( p^2 k^2 - (pk)^2 \bigg) \, \tau_4
                ~ + ~ 3 \, k^2 \, \tau_5
                ~ + ~  \bigg(\frac{3}{2} \, k^4 + 2 \, k^2 p^2 + 6 \, k^2 (pk) + 4 \, (pk)^2 \bigg) \, \tau_7  ,
                \label{FFtau4-Omega2}
 \\
 \Omega_3 & = & 
             \frac{1}{4} \, \mbox{Tr} \Big( \sigma_{\alpha\mu} \, p^\alpha ~ \Gamma^\mu  \Big) 
                ~ = 
                \bigg( p^2 k^2 - (pk)^2 \bigg) (pk) \,  \tau_4
                +  3 \, (pk) \,  \tau_5
                +  \bigg( 2 \, p^2 k^2  + \frac{3}{2} \, k^2 (pk) + 6 \, p^2 (pk) + 4 \, (pk)^2 \bigg)  \tau_7  .
                                \label{FFtau4-Omega3}
\end{eqnarray}
Indeed, it comes that
\begin{eqnarray}
\tau_4 & = & 
  \frac{1}{2 \, \Big( k^2 + 2 \, (pk) \Big) \, \Delta^2} \bigg(
         - \, 3 \, \Big(\, k^2 + 2 \, (pk)\Big) \, \Omega_1
         ~ + ~ \Omega_2
         ~ - ~ 2 \, \Delta \, \Omega_3 \bigg) \ ,
\\
\tau_5 & = & 
         \frac{1}{4 \, \Delta} \bigg( 
                 2 \, \Omega_1 ~  + ~  \Big( 2 \, p^2 + (pk) \Big) \, \Omega_2 
                 ~ - ~  \Big( k^2  + 2 \, (pk) \Big) \, \Omega_3\bigg) \ ,
\\
\tau_7 & = & 
   - \, \frac{1}{2 \, \Big( k^2 + 2 \, (pk) \Big)\,  \Delta} \bigg(
                        (pk) \, \Omega_2   ~ - ~ k^2 \, \Omega_3 \bigg) \ .
\end{eqnarray}
Again, as $\tau_{1, \, 2, \, 3, \, 6}$ the form factors are all  proportional to $1 / \Delta$. Furthermore, for a tree level type of vertex one has that
$\Omega_i = 0$ and, therefore, the $\tau_{4, \, 5, \, 7}$  vanish.

\section{The Vertex DSE, the form factors and  approximations \label{Sec:softphotonlimit}}

The r.h.s. of this Dyson-Schwinger equation (\ref{DSER-vertex}), has besides a ``tree-level'' type of contribution, two terms requiring
momentum integration.
In one of these two terms the full vertex appears twice, with different momenta setups, while in the other a contribution of the two-photon-two-fermion 
one-particle irreducible Green function has to be evaluated. The first term, that requires only $\Gamma^\mu$ itself, needs the following vertices 
\begin{equation}
\Gamma^\mu ( U, \, -U-k; \, k) \qquad\mbox{ and }\qquad 
\Gamma^{\zeta^\prime} (V, \, -V-q; \, q) \ , 
\end{equation}
with the momenta $U = p -q$, $V = p +k -q = U + k$ and where $q$ is the loop momentum that appears in the integration. 
The full vertex $\Gamma^\mu$ requires twelve form factors. The consideration of its full structure in the integral equation results in a lengthy expression 
that is hard to understand and disentangle.  To study the vertex DSE itself and to be able to solve it, approximations are welcome. 
The considerations below apply to the analysis only of the r.h.s. of the vertex Dyson-Schwinger equation, i.e. to the terms inside the momentum integral. 
The approximations to be discussed are inspired in the perturbative program for QED.

To solve the vertex integral equation, in its r.h.s., the contributions of the transverse form factors $\tau_i$ are disregarded and only 
the longitudinal form factors are considered. The $\lambda_i$'s are the solutions given in Eqs (\ref{EQ:L1}) to (\ref{EQ:L4}) of the vertex WTI. 
By taking into account all the longitudinal form factors one goes beyond the conventional perturbation theory, whose starting point sets
$\lambda_1 = 1$ and ignores $\lambda_{2,3}$. 
Despite this approximation, the r.h.s. of the vertex equation still has a large number of terms and, to simplify further, the two-photon-two-fermion term
will not be taken into account.
In perturbation theory, the two-photon-two-fermion 1PI Green function vanishes at tree level and gets a first contribution from one-loop diagrams, that is of order $g^4$.
See \cite{Oliveira:2022bar} for the DSE that determines $\Gamma^{\mu\nu}$ and its diagrammatic representation.
From this point of view, it seems reasonable to ignore $\Gamma^{\mu\nu}$ in a first approximation. 
Despite the approximations, they still result in large expressions that are difficult to handle and further simplifications will be considered to compute the $\tau_i$'s.
To reduce further the number of terms in the r.h.s. of the equation we take, only in its r.h.s., the chiral limit defined by setting  $m = B = 0$. 
Then, to be consistent with the WTI (\ref{EQ:L3}), in the r.h.s., we also ignore the contribution of the $\lambda_3$ form factor. 
According to one-loop perturbation theory, in the chiral limit only a subset of the transverse form factors has to be considered. Our analysis of the vertex integral
equation finds the some set of non-vanishing transverse form factors as in the perturbative solution.

\textcolor{blue}{The approximations discussed to solve the vertex equation do not consider the full vertex in the r.h.s. of the equation and, therefore,  
can compromise the multiplicative renormalizability of the theory. Despite the approximations, it turns out that the solution for $\Gamma^\mu$ result in
transverse form factors that do not all vanish. 
The presence of the $\tau_i$'s is crucial to recover such an important property for QED, see e.g. \cite{Curtis:1990zs,Kizilersu:2009kg} and references therein. Although we are not able to proof
that the approximations introduced do not lead to a violation of the multiplicative renormalizability, the presence of the various $\tau_i$ in the solution seems to
suggest that the violation of multiplicative renormalizability should be small.}

\textcolor{blue}{The results in this section can also be viewed in two ways. On one hand, the approximations in the r.h.s. of the vertex equation
identify the dominant tensor structures in $\Gamma^\mu$ that are expected to be dominant, i.e. identify the tensor structures that can be disregarded in a first approximation.
On the other hand, assuming a power series solution, in terms of the coupling constant, of the vertex equation, it suggests that deviations should occur at larger order in $g$ 
and, therefore, possible violations of multiplicative renormalizability should take place at $\mathcal{O}(g^4)$ or at higher order.
The perturbative analysis performed in Sec. \ref{Sec:PertFFactors} solves exactly the vertex equation up to $g^2$ and reproduces the gross features of the
standard perturbative vertex calculation.
}

Let us proceed with the analysis of vertex DSE to compute the transverse form factors using the above approximations in the evaluation of its r.h.s. The calculation
relies on the main results of Sec. \ref{SecTransverseFF}.
In the analysis of the Dirac structure of the vertex Dyson-Schwinger integral equation our focus goes to the terms under the integral. 
Within the approximations discussed, this terms reads, recall that we are disregarding the contribution of $\Gamma^{\mu\nu}$,
\begin{eqnarray}
& & 
\hspace{-0.25cm}
D_{\zeta\zeta^\prime}(q)  ~ \gamma^\zeta ~ S(p-q) ~  {\Gamma}^\mu ( p - q, \, -p -k + q; \, k) ~ 
                                S(p+k-q) ~  {\Gamma}^{\zeta^\prime} (p +k-q, \, -p-k; q) ~  =
                                \nonumber \\
& & \quad
= - D(q^2) ~ \left[    \gamma^\zeta ~ S(p-q) ~  {\Gamma}^\mu ( p - q, \, -p -k + q; \, k) ~ 
                                S(p+k-q) ~  {\Gamma}_{\zeta} (p +k-q, \, -p-k; q) \right]
                                \nonumber \\
& & \quad\quad
- \, \frac{1}{q^2} \left( \frac{\xi}{q^2} - D(q^2) \right)  
  \left[ \slashed{q} \, S(p-q) \,  {\Gamma}^\mu ( p - q, \, -p -k + q; \, k) \,
                                S(p+k-q)  \Big( q^\zeta {\Gamma}_{\zeta} (p +k-q, \, -p-k; q\Big) \right]
                                \nonumber \\
& & \quad
= - D(q^2) \, \widetilde{\Gamma}^\mu_{(D)} - \, \frac{1}{q^2} \left( \frac{\xi}{q^2} - D(q^2) \right)   \widetilde{\Gamma}^\mu_{(\xi)} \ .
\label{VertexEq-Sep-Int}
\end{eqnarray}
A global factor of
\begin{equation}
i \, g^2 \,  Z_2 \, \int \frac{d^4q}{(2 \, \pi)^4} 
\end{equation}
is omitted to simplify the notation.
Before proceeding, we remind that in QED the combination $g^2 \, D(q^2)$ is renormalization scale independent and can be seen 
as an effective charge. For the Landau gauge ($\xi = 0)$, this effective charge multiplies all terms in the vertex Dyson-Schwinger equation.
For other gauges the role of the effective charge is not so obvious. In this sense the Landau gauge is a special gauge. Note, however, that
in the quenched approximation where $D(q^2) = 1/q^2$, the Feynman gauge ($\xi = 1$) shares this same property.

\subsection{Getting $\tau_1$}

The evaluation of this form factor was discussed in Sec. \ref{SubSecTau1} and is summarized in Eq. (\ref{tau1-fromvertex-Mink}).
The evaluation of the traces
\begin{equation}
  \mbox{Tr} \left( p_\mu \, \widetilde{\Gamma}^\mu_{(D)} \right) \qquad\mbox{ and }\qquad
  \mbox{Tr} \left( p_\mu \, \widetilde{\Gamma}^\mu_{(\xi)} \right)
\end{equation}
give terms that are proportional to $\lambda_3$ and if this form factor vanish, then 
\begin{equation}
  \tau_1  \Big((p+k)^2, \, p^2, \, k^2\Big)=   ~ \frac{2 \, p^2 + (pk)}{ \Delta \,  \Big( k^2 + 2 \, (pk)  \Big) } 
             ~ \Big( B((p+k)^2) - B( p^2) \Big) 
\  .
\label{tau1-fromvertex-chiral}
\end{equation}
For a constant $B(p^2)$ or in the chiral limit, i.e. for massless fermions, this form factor vanish.

\subsection{Getting $\tau_8$}

For the evaluation of $\tau_8$ we have to consider Eq. (\ref{GetTau8}) and contract it with a Levi-Civita tensor arriving at
\begin{equation}
 {\tau_8} ~~ \epsilon_{\zeta\mu\alpha\beta} ~p^\alpha k^\beta   ~~ {\epsilon }^{\zeta  \mu  \alpha^\prime\beta^\prime} ~~  {p}_{\alpha^\prime} \, {k}_{\beta^\prime} ~ = ~
 2 ~ {\tau_8} ~ \Big( (pk)^2 -  p^2 k^2 \Big)  ~= ~ 
 \frac{i}{4} \mbox{Tr} \Big( \gamma^\zeta \, \gamma_5 \, \Gamma^\mu \Big) ~~ \epsilon_{\zeta\mu\alpha\beta} ~p^\alpha k^\beta \ .
\end{equation}
The corresponding contractions with $\widetilde{\Gamma}^\mu_{(D)}$ and $\widetilde{\Gamma}^\mu_{(\xi)}$ produce lengthy expressions that require
the products $\lambda_1 \lambda_1$, $\lambda_1 \lambda_2$ and $\lambda_2 \lambda_2$, with the two form factors evaluated at different kinematical points.
These expressions simplify in the limit of $\lambda_2 = 0$, that corresponds to a $A(p^2)$ that is essentially constant. In this simplified version the traces read
\begin{equation}
\widetilde{\Gamma}^\mu_{(D)} \quad \to \quad 4  ~ \lambda_1 \Big( \, (U+k)^2, \, U^2, \, k^2 \, \Big) ~ \lambda_1  \Big(\, (V+q)^2, \, V^2, q^2 \, \Big) ~ \Big[ (kU) (pV) - (kV) (pU) \Big]
\end{equation}
and
\begin{eqnarray}
& & 
\widetilde{\Gamma}^\mu_{(\xi)} \quad \to \quad 2 ~  \lambda_1 \Big( \, (U+k)^2, \, U^2, \, k^2 \, \Big) ~ \lambda_1  \Big(\, (V+q)^2, \, V^2, q^2 \, \Big) 
\nonumber \\
& & \qquad\qquad\qquad
 \Bigg[
   (qU) \bigg( \, (kq) (pV) - (kV) (pq) \, \bigg)
   \, + \, 
   (qV) \bigg( \, (kU) (pq) - (kq) (pU) \, \bigg)
   \Bigg] \ .
\end{eqnarray}
From these results one can build the renormalized expression 
\begin{align}
 \tau_8 ( (p+k)^2, p^2, k^2 ) & =  ~ \frac{i \, g^2 \,  Z_2}{ \Delta  } \, \int \frac{d^4q}{(2 \, \pi)^4} ~ \lambda_1 \Big( \, (U+k)^2, \, U^2, \, k^2 \, \Big) ~ \lambda_1  \Big(\, (V+q)^2, \, V^2, q^2 \, \Big) 
      ~ 
          \nonumber \\
    & \qquad
    \frac{A (U^2)}{ A^2 (U^2) \, U^2 - B^2 (U^2)}~ \frac{A (V^2)}{A^2 (V^2) \, V^2 - B^2 (V^2)}
     ~ \Bigg\{
    2 \,  D(q^2) \, \bigg[ (kU) (pV) - (kV) (pU) \bigg] \, 
      \nonumber \\
      &  \quad\qquad ~
      + \, \frac{1}{q^2} \left( \frac{\xi}{q^2} - D(q^2) \right)  \, \Bigg[
   (qU) \bigg( \, (kq) (pV) - (kV) (pq) \, \bigg)
   \, + \, 
   (qV) \bigg( \, (kU) (pq) - (kq) (pU) \, \bigg)
   \Bigg]  \Bigg\}
   \label{Sol:tau8}
\end{align}
that will be explored later in a perturbative type of approach.

\subsection{Getting $\tau_2$, $\tau_3$ and $\tau_6$}

The starting point for the computation of $\tau_2$, $\tau_3$ and $\tau_6$  are the expressions for the $\Lambda_i$'s derived in Sec. \ref{SecGetTau236}, see
Eqs (\ref{FFtau2:Exact}), (\ref{FFtau3:Exact}) and (\ref{FFtau6:Exact}).
The complete expressions for each of these quantities is, once more, rather lengthy and, as in the calculation of $\tau_8$, requires
the products $\lambda_i \lambda_j$ at different kinematical points. A simplified version can be obtain ignoring $\lambda_2$. For the trace appearing in $\Lambda_1$
it follows that
\begin{equation}
\widetilde{\Gamma}^\mu_{(D)} \quad \to \quad 4  ~ \lambda_1 \Big( \, (U+k)^2, \, U^2, \, k^2 \, \Big) ~ \lambda_1  \Big(\, (V+q)^2, \, V^2, q^2 \, \Big) ~  (UV)
\end{equation}
and
\begin{eqnarray}
& & 
\widetilde{\Gamma}^\mu_{(\xi)} \quad \to \quad 4 ~  \lambda_1 \Big( \, (U+k)^2, \, U^2, \, k^2 \, \Big) ~ \lambda_1  \Big(\, (V+q)^2, \, V^2, q^2 \, \Big) 
    ~ (qU)  ~ (qV) \ .
 \end{eqnarray}
For the trace appearing in $\Lambda_2$ and $\Lambda_3$ there are no contribution coming from $\widetilde{\Gamma}^\mu_{(D)}$ and $\widetilde{\Gamma}^\mu_{(\xi)}$, 
even if $\lambda_2$ is taken into account. 
Then, for $\Lambda_2$ and $\Lambda_3$ the traces in their definitions vanish in the chiral limit, i.e. for $B = 0$ and $\lambda_3 = 0$ and they are given by
\begin{eqnarray}
\Lambda_2 & = &   -  (pk) \, \lambda_1 \Big( (p+k)^2, p^2, k^2\Big)  - \Big(  2 \, p^2 + (p k) \Big) \Big( k^2 + 2  (p k) \Big) \, \lambda_2 \Big( (p+k)^2, p^2, k^2\Big)   ,
 \\      
\Lambda_3 & = &   -  p^2 \, \lambda_1 \Big( (p+k)^2, p^2, k^2\Big) -  \Big( 2 \, p^2 + (pk) \Big)^2 \, \lambda_2 \Big( (p+k)^2, p^2, k^2\Big)
\end{eqnarray}
and with the help of Eqs (\ref{EQ:L1}) and (\ref{EQ:L2}) they can be rewritten in terms of the fermion propagator function $A(p^2)$. On the other hand
\begin{align}
\Lambda_1 & = 4 \, Z_2   ~ - ~ 4 \, i \, g^2 \,  Z_2 \, \int \frac{d^4q}{(2 \, \pi)^4} ~ \lambda_1 \Big( \, (U+k)^2, \, U^2, \, k^2 \, \Big) ~ \lambda_1  \Big(\, (V+q)^2, \, V^2, q^2 \, \Big) ~ 
 \nonumber 
 \\
 & \hspace{2cm}
  \frac{A(U^2)}{A^2(U^2) \, U^2 - B^2(U^2)} ~ \frac{A(V^2)}{A^2(V^2) \, V^2 - B^2(V^2)}
~  \Bigg\{ D(q^2) ~(UV) ~ +  ~ \frac{1}{q^2} \left( \frac{\xi}{q^2} - D(q^2) \right)   ~ (qU)  ~ (qV) 
 \Bigg\} \ .
 \label{Eq:Lambda1Approx}
\end{align}
Then, it follows from Eqs (\ref{FFtau2:Exact}) to (\ref{FFtau6:Exact}) that the transverse form factors are given by
\begin{eqnarray}
\tau_2 & = &
- \,\frac{1}{4 \, \Delta} \, \Lambda_1
      ~ - ~ \frac{3}{4  \, \Delta} \Bigg( \lambda_1 + 2 \, \Big( 2 \, p^2 + (pk) \Big) \lambda_2 \Bigg) 
 ~ = ~
- \,\frac{1}{4 \, \Delta} ~ \Bigg\{  \Lambda_1 ~ + ~ 3 ~ \bigg[ \lambda_1 + 2 \, \Big( 2 \, p^2 + (pk) \Big) \lambda_2 \bigg] \Bigg\}
 \ ,     \label{FFtau2:ExactApprox}
\\
\tau_3 & = & 
        \frac{1}{8 \, \Delta}  \Bigg\{ 
           \bigg(  k^2 + 4 \,  p^2 + 4 \,  (pk)  \bigg) \, \Lambda_1 
         +  k^2 ~ \Bigg(   \bigg[ 3 + 4 \, \frac{p^2}{k^2} + 8 \, \frac{(pk)}{k^2} \bigg] \, \lambda_1 
                                                         + \Big( 2 \, p^2 + (pk) \Big) \bigg[ 2 + 8 \, \frac{p^2}{k^2} + 8 \, \frac{(pk)}{k^2} \bigg] \, \lambda_2 \Bigg)  
                                                         \Bigg\}        ,
     \label{FFtau3:ExactApprox}
\\
\tau_6 & = &
 - \, \frac{1}{8 \, \Delta} \Bigg\{
\bigg(  k^2 + 2 \, (pk)  \bigg)\,  \Lambda_1
+   k^2 \Bigg(  \bigg[ 3 + 2 \, \frac{(pk)}{k^2} \bigg] \, \lambda_1 +  \Big( 2 \, p^2 + (pk) \Big) \bigg[ 2 + 4 \, \frac{(pk)}{k^2} \bigg] \, \lambda_2 \Bigg) 
  \Bigg\}   \ .
     \label{FFtau6:ExactApprox}
\end{eqnarray}
The form factors $\lambda_1$ and $\lambda_2$ are related to the fermion propagator form factor $A(p^2)$
and the above expressions call be rewritten in terms of $A(p^2)$.
The form factors in Eqs (\ref{FFtau2:ExactApprox}) to (\ref{FFtau6:ExactApprox}) are evaluated at the kinematical point $ ( (p+k)^2, p^2, k^2 )$. 
The above expressions for $\tau_3$ and for $\tau_6$ imply that, within the approximations considered, they comply with the equality
\begin{align}
\tau_3 + \tau_6 & = 
   \frac{1}{4 \, \Delta} ~
           \bigg(  2 \,  p^2 +  (pk)  \bigg) \, \Lambda_1 
~ +  ~ \frac{1}{4 \, \Delta}  ~ \Bigg(   \bigg( 2 \, p^2 + 3 \, (pk) \bigg) \, \lambda_1 
                                                         + 2 \, \bigg( 2 \, p^2 + (pk) \bigg)^2  \, \lambda_2 \Bigg)  \ . 
    \nonumber \\
 & = 
   \frac{ 2 \,  p^2 +  (pk) }{4 \, \Delta} ~ \Bigg\{
          ~ \Lambda_1 
~ +  ~   \frac{ 2 \, p^2 + 3 \, (pk) }{ 2 \,  p^2 +  (pk) } ~ \lambda_1 
                                                         + 2 \, \bigg( 2 \, p^2 + (pk) \bigg)  ~ \lambda_2 \Bigg\}  \ .                                                              
\end{align}
This particular combination of the two form factors appears in the expression of the full vertex, see Eq. (\ref{FullVertex-Dirac-main}), and the last equation can be
used to simplify the writing of the vertex.

\subsection{Getting $\tau_4$, $\tau_5$ and $\tau_7$}

Of the full set of transverse form factors, it remains to discuss the evaluation of $\tau_4$, $\tau_5$ and $\tau_7$ from the vertex equation. As a first step, the computation of
the $\Omega_i$'s defined in Eqs (\ref{FFtau4-Omega1}) to (\ref{FFtau4-Omega3}) is required. In the chiral limit, as defined previously, it turns out that
all the $\Omega_i$ vanish and, therefore, in this limit $\tau_4 = \tau_5 = \tau_7 = 0$. In one-loop perturbation theory these form factors are proportional to the fermion
mass and vanish in the chiral limit.

\subsection{On the transverse form factors and the chiral limit of the vertex equation}

Our discussion for the transverse form factors $\tau_i$ using the vertex equation rely on an exact solution of the vertex WTI, given in Eqs (\ref{EQ:L1}) to (\ref{EQ:L4}).
In the r.h.s. of the equation, the chiral limit inside the integral is taken, implying to set $m = B = 0$ and to take $\lambda_3 = 0$ to be consistent with the solution of the vertex WTI. 
Further, the contribution of the two-photon-two-fermion Green function to the solution is not taken into account. 
To arrive at manageable expressions the contribution of $\lambda_2$ is, sometimes, ignored. Despite the approximations considered, the calculation using the
vertex DSE confirms the estimation of the one-loop perturbative calculation \cite{Kizilersu:1995iz} for the $\tau_i$ that, for massless fermions,
the non-vanishing form factors to be considered are $\tau_2$, $\tau_3$, $\tau_6$ and $\tau_8$. 
Moreover, the derived expressions show that the combination of momenta $\Delta$, see Eq. (\ref{FFDelta}), defines a momentum scale for the transverse form factors
as in the perturbative calculation. 

\subsection{On the asymptotic scaling of the transverse form factors}

The results just derived for the transverse form factor allow to describe their asymptotic behaviour, both in the limit where the incoming fermion momentum $p$ or 
the incoming photon momentum $k$ are large. The combination of momenta $\Delta$ defined in Eq. (\ref{FFDelta}) defines a scale for all the transverse form factors. 
To study the asymptotic regime it is convenient to write 
\begin{equation}
\Delta = p^2 k^2 \left( 1 - \frac{(pk)^2}{p^2 k^2} \right) =  p^2 k^2 \left( 1 - \cos^2\theta_{pk} \right)  \ .
\end{equation}
In what concerns power counting, in the large $p$ and/or $k$ limit one can set $\Delta \approx p^2 \, k^2$. Then,  naive power counting gives
\begin{equation}
  \tau_1  \Big(k^2, \, p^2, \, k^2\Big) \approx   ~ \frac{(pk)}{ p^2 k^4 } 
             ~ \Big( B(k^2) - B( p^2) \Big)  \ .
\label{tau1-fromvertex-chiral-asympk}
\end{equation}

 The corresponding analysis for $\tau_8 ( k^2, p^2, k^2 )$ is more involved due to the momentum integration that appears on its definition. 
A full answer requires being able to perform the integrations to understand the leading asymptotic behaviour. This can certainly be done in perturbation
theory and it will be considered in Sec. \ref{Sec:PertFFactors}. 

For the study of the asymptotic behaviour of $\tau_{2,3,6}$ the longitudinal form factors $\lambda_2$ and $\lambda_3$ are sub-leading relative to $\lambda_1$ and, therefore, 
the leading term is associated with the later longitudinal form factor. At large $k$ it comes that
\begin{eqnarray}
\tau_2 ( k^2, p^2, k^2 ) & \approx &
- \,\frac{Z_2}{p^2 k^2} 
      ~ - ~ \frac{3}{8  \,p^2 k^2} \Bigg( A(k^2) + A(p^2)  \Bigg) \ ,
     \label{FFtau2:ExactApprox-asymptk}
\\
\tau_3 ( k^2, p^2, k^2 ) & \approx & 
\frac{Z_2}{2 \, p^2} 
~ +   \frac{3}{16 \, p^2}  \Bigg( A(k^2) + A(p^2)  \Bigg) ,
     \label{FFtau3:ExactApprox-asymptk}
\\
\tau_6 ( k^2, p^2, k^2 ) & \approx &
 - \, \frac{Z_2}{2 \, p^2 } 
~ -   \frac{3}{16 \, p^2} \Bigg( A(k^2) + A(p^2)  \Bigg)
     \label{FFtau6:ExactApprox-asymptk}
\end{eqnarray}
and all these form factors, together with the sum $\tau_3 + \tau_6$, become irrelevant at large momenta. 

Naive power counting suggest that, in the large photon momentum limit, the vertex is dominated by the longitudinal form factors. In this case, for large $k$ the vertex reads
\begin{eqnarray}
{\Gamma}^\mu (p, \, -p -k; \, k)        & \approx &
      - {k}^{\mu }  ~  {\lambda_3} ( k^2, \, p^2, k^2) 
    ~ + ~ 
     {\gamma }^{\mu }  ~ {\lambda_1}  ( k^2, \, p^2, k^2)
     ~ + ~ \slashed{k}  \,  {k}^{\mu }  ~ {\lambda_2}  ( k^2, \, p^2, k^2)
     \ ,
     \label{FullVertex-Dirac-main-asymptk}
\end{eqnarray}
where
\begin{align}
{\lambda_1}  ( k^2, \, p^2, k^2) & = \frac{1}{2} \Big( A(k^2) + A(p^2) \Big) \ , \\
{\lambda_2}  ( k^2, \, p^2, k^2) &= \frac{1}{2} \frac{1}{k^2 - p^2} \Big( A(k^2) - A(p^2) \Big) \ ,  \\
{\lambda_3}  ( k^2, \, p^2, k^2) &=  \frac{1}{k^2 - p^2} \Big( B(k^2) - B(p^2) \Big) \ .
\end{align}
On the other hand, for large $p$, the photon-fermion vertex is given by
\begin{eqnarray}
{\Gamma}^\mu (p, \, -p -k; \, k)   & = & 
     - \, 2 \, {p}^{\mu } ~  {\lambda_3} 
    ~ + ~ 
     {\gamma }^{\mu } \, {\lambda_1} 
     ~ + ~ 4 \, \slashed{p} \,  {p}^{\mu }  \, {\lambda_2} 
     \ ,
     \label{FullVertex-Dirac-main-asymptp}
\end{eqnarray}
where the longitudinal form factors are
\begin{equation}
   \lambda_1 \Big(p^2, \, p^2, \, k^2\Big) = A(p^2) \, \qquad
   \lambda_2 \Big(p^2, \, p^2, \, k^2\Big) = \frac{1}{2} A^\prime(p^2) \, \qquad\mbox{ and }\qquad
   \lambda_3 \Big(p^2, \, p^2, \, k^2\Big) = B^\prime(p^2) \ .
\end{equation}   
If the terms with derivatives of $A$ and/or $B$ are sub-leading, then the above expression can be further simplified.

\section{Perturbative estimation of the transverse form factors \label{Sec:PertFFactors}}

The above expressions for the transverse form factors can be evaluated with perturbation theory.  This calculation requires a discussion of the renormalization in QED.
The renormalization constants can be computed requiring
\begin{equation}
A(\mu^2_F) = 1, \qquad\mbox{ and }\qquad B(\mu^2_F) = m^{(phys)}
\label{Eq:renormalization}
\end{equation}
where $\mu_F$ is the renormalization mass scale and $m^{(phys)}$ the physical fermion mass. The bare functions, evaluated using dimensional regularization and
in Minkowski spacetime, that determine the inverse of the fermion propagator were computed to one-loop order in the coupling constant in \cite{Kizilersu:1995iz} and are given by
\begin{align}
   A^{(0)}(p^2) & = 1 + \frac{ \alpha \, \xi}{4 \, \pi} \Bigg[ C \, \mu^\epsilon + \left( 1 + \frac{m^2_0}{p^2} \right) \Big( 1 - L(p^2) \Big) \Bigg] \label{Eq:PertA}\\
   B^{(0)}(p^2) & = m_0 ~\Bigg\{  1 +  \frac{ \alpha \, \xi}{4 \, \pi} \Bigg[  2 + C \, \mu^\epsilon - L(p^2)  \Bigg] 
   +  \frac{ \alpha}{\pi} \Bigg[  1 + \frac{3}{4} \, C \, \mu^\epsilon  - \frac{3}{4} \,  L(p^2)  \Bigg] \Bigg\} \label{Eq:PertB}
\end{align}
where $m_0$ is the bare fermion mass, the spacetime dimension is $D = 4 - \epsilon$, $\mu$ is the usual mass parameter introduced to ensure that the coupling $\alpha$ remains dimensionless for any $D$ and
\begin{align}
C & =  \frac{2}{\epsilon} - \gamma_E + \ln 4 \, \pi + \ln\frac{m^2}{\mu^2} \ , \\
L(p^2) & = \left( 1 - \frac{m^2}{p^2} \right) \, \ln \left( 1 - \frac{p^2}{m^2} \right) \  . 
\end{align}
To first order in the coupling constant the bare coupling $\alpha_0$ is undistinguishable from the renormalized $\alpha$.
Further, the expressions for the form factors are all multiplied by $g^2$ and, for the level of precision under discussion, 
one can rely on the tree level gauge boson propagator.
It follows from the renormalization conditions (\ref{Eq:renormalization}) that the corresponding renormalized functions are
\begin{widetext}
\begin{align}
   A(p^2) & = 1 + \frac{ \alpha \, \xi}{4 \, \pi} \Bigg[   L(\mu^2_F)  - L(p^2)  + \frac{m^2}{p^2}  \Big( 1 - L(p^2) \Big) -  \frac{m^2}{\mu^2_F}  \Big( 1 - L(\mu^2_F) \Big) \Bigg]
   \label{A-1Loop-Ren}  \\
   B(p^2) & = m ~\Bigg\{  1 +  \frac{ \alpha}{4 \, \pi} \Big( 3 + \xi \Big) \Bigg[   L(\mu^2_F)   -   L(p^2)  \Bigg] \Bigg\} \ ,\label{B-1Loop-Ren}
\end{align}
\end{widetext}
where $m$ is  the physical fermion mass. Further, the fermion field and the mass renormalization constants are 
\begin{align}
   Z_2 & = 1 - \frac{ \alpha \, \xi}{4 \, \pi} \Bigg[ C \, \mu^\epsilon + \left( 1 + \frac{m^2}{\mu^2_F} \right) \Big( 1 - L(\mu^2_F) \Big) \Bigg]  \label{Def:Z2Pert} \\
   Z_0 & = 1 - \frac{ \alpha \, \xi}{4 \, \pi} \Bigg[ 2 + C \, \mu^\epsilon - L(\mu^2_F)  \Bigg]
                    - \frac{ \alpha}{\pi} \Bigg[ 1 + \frac{3}{4} \, C \, \mu^\epsilon - \frac{3}{4} L(\mu^2_F) \Bigg] \ ,
\end{align}
respectively.  A straightforward calculation show that the perturbative solution for the transverse vertex in \cite{Kizilersu:1995iz}  reproduces the solution
of the WTI. 

The perturbative estimation of $\tau_1$, i.e. of Eq. (\ref{tau1-fromvertex-chiral}), is 
\begin{align}
  \tau_1  \Big((p+k)^2, \, p^2, \, k^2\Big) &  =   ~    m \, \frac{\alpha}{4 \, \pi} \, (3 + \xi) ~
\frac{2 \, p^2 + (pk)}{ \Big( p^2 k^2 - (pk)^2 \Big) \,  \Big( k^2 + 2 \, (pk)  \Big) }  
  \\
  & \qquad
       ~ \Bigg\{  \ln \left( \frac{ m^2 - p^2}{m^2 - (p+k)^2} \right) ~ + ~ \frac{m^2}{(p+k)^2} \, \ln \left( 1 - \frac{(p+k)^2}{m^2} \right)
                   ~ - ~ \frac{m^2}{p^2} \, \ln \left( 1 - \frac{p^2}{m^2} \right) \Bigg\}
\  .
\label{Pert:tau1}
\end{align}
In the small momentum limit where $p^2, \, k^2 \lesssim m^2$ this form factor is given by
\begin{align}
  \tau_1  \Big((p+k)^2, \, p^2, \, k^2\Big) &  \approx   ~   \frac{1}{m} ~  \frac{\alpha \, ( 3 + \xi) }{8 \, \pi} ~  \frac{2\, p^2 + (pk) }{ p^2 k^2 - (pk)^2 }  \ .
\end{align}
The divergence at low momenta is not a problem for the DSE as it is the combination $k^2 \, \tau_1$,
multiplied by other factors of momenta coming from the integration measure, that appears
in the integral equations, see Eqs (\ref{DSE-Fermion-scalar}), (\ref{DSE-Fermion-vector}) and (\ref{Eq:PhotonEqBasis}).
In the large momentum limit, the expression for $\tau_1$ simplifies but its exact expression depends on the relative importance of the momenta $p^2$ and $k^2$.
The above expression share with the one-loop perturbative solution the common global factor
\begin{align}
 \frac{\alpha}{4 \, \pi} ~
\frac{ (3 + \xi) }{ p^2 k^2 - (pk)^2 } ~ m \ .
\end{align}
The perturbative solution can be found in  \cite{Kizilersu:1995iz} and is written in terms of Spence functions.

Let us now compute $\tau_8$ within the approximation scheme considered so far. Its expression, see Eq. (\ref{Sol:tau8}), is given by
\begin{align}
 \tau_8 ( (p+k)^2, p^2, k^2 ) & =  ~ \frac{i \, g^2 \,  Z_2}{ \Delta  } \, \int \frac{d^4q}{(2 \, \pi)^4} ~ 
    \frac{1}{q^2} ~\frac{1}{ U^2 - m^2}~ \frac{1}{V^2 - m^2}
     ~ \Bigg\{
    2 \,   \bigg[ (kU) (pV) - (kV) (pU) \bigg] \, 
      \nonumber \\
      &  \quad\qquad ~
      + \, \frac{1}{q^2}  \left( \xi - 1 \right)  \, \Bigg[
   (qU) \bigg( \, (kq) (pV) - (kV) (pq) \, \bigg)
   \, + \, 
   (qV) \bigg( \, (kU) (pq) - (kq) (pU) \, \bigg)
   \Bigg]  \Bigg\}
    \nonumber \\
   & =  ~ \frac{i \, g^2 \,  Z_2}{ \Delta  } \, \int \frac{d^4q}{(2 \, \pi)^4} ~ 
    \frac{1}{q^2} ~\frac{1}{ (p-q)^2 - m^2}~ \frac{1}{(p + k  -q)^2 - m^2}
      \nonumber \\
      &  \quad\qquad ~
     ~ \Bigg\{
    2 \,   \bigg[  (pk)^2 ~ - ~ k^2 \, p^2  ~ + ~ k^2 \, (pq) ~ - ~ (pk) \, (kq)  \bigg] \, 
      \nonumber \\
      &  \qquad\qquad\qquad ~
      + \, \frac{1}{q^2}  \left( \xi - 1 \right)  \, \Bigg[  2 \, (pk)  \, (pq)\,  (kq) ~  - ~ k^2 \, (pq)^2  ~ - ~ p^2 \, (kq)^2  
                                                                             \nonumber \\
                                                                             & \hspace{6cm}
                                                                             ~ + ~ q^2 \,  \Big( k^2 \, (pq)    ~ - ~  (pk) \, (kq) \Big)
 \Bigg]  \Bigg\} 
 \ .
\end{align}
The momentum integrations can be performed within dimensional regularization,  after the introduction of Feynman parameters.
The relevant momentum integrations are summarized in App. \ref{App:MomInt}. After a straightforward algebra, this form factors reads
\begin{align}
 \tau_8 ( (p+k)^2, p^2, k^2 ) ~ =  
     ~ -  \, \frac{\alpha}{4 \, \pi}  ~ \Bigg\{ &
     ~ \, \int^1_0dx \int^x_0dy ~ \frac{x-2}{\mathcal{M}^2 - K^2} ~ + ~ \, \int^1_0dx \int^x_0dy ~ \frac{y \, \mathcal{M}^2}{\left( \mathcal{M}^2 - K^2 \right)^2}
     \nonumber \\
     & ~+~
     \xi \Bigg[
          ~ \, \int^1_0dx \int^x_0dy ~ \frac{x}{\mathcal{M}^2 - K^2} ~ + ~ \, \int^1_0dx \int^x_0dy ~ \frac{y \, \mathcal{M}^2}{\left( \mathcal{M}^2 - K^2 \right)^2}
     \Bigg]
      \Bigg\} 
 \ ,
\end{align}
where $K = x \, p + y k$ and  $\mathcal{M}^2 = x\, ( p^2 - m^2) + y ( k^2 + 2 \, (pk) ) + i \, \epsilon$. For large photon momentum,
$K^2  = y^2 k^2$ and  $\mathcal{M}^2 = k^2 ( y + i \, \epsilon/k^2 )$ and it follows that
\begin{align}
 \tau_8 ( (p+k)^2, p^2, k^2 ) ~ =  
     ~ -  \, \frac{\alpha}{4 \, \pi}  ~ \frac{1}{k^2} \Bigg\{ &
     ~ \,  \int^1_0dx \int^x_0dy ~ \frac{x-2}{y(1-y) + i \epsilon/k^2} ~ + ~  \int^1_0dx \int^x_0dy ~ \frac{y^2}{\left(y(1-y) + i \epsilon/k^2\right)^2}
     \nonumber \\
     & ~+~
     \xi \Bigg[
          ~ \, \int^1_0dx \int^x_0dy ~ \frac{x}{y(1-y) + i \epsilon/k^2} ~ + ~ \, \int^1_0dx \int^x_0dy ~ \frac{y^2 }{\left( y(1-y) + i \epsilon/k^2 \right)^2}
     \Bigg]
      \Bigg\} 
 \ ,
\end{align}
while for large fermion momentum
\begin{align}
 \tau_8 ( (p+k)^2, p^2, k^2 ) ~ =  
     ~ -  \, \frac{\alpha}{4 \, \pi}  ~ \frac{1}{p^2} \, \Bigg\{ &
     ~ \, \int^1_0dx  ~ \frac{x(x-2)}{x(1-x) + i \epsilon/p^2} ~ + ~ \, \frac{1}{2} \, \int^1_0dx  ~ \frac{x^3}{\left( x(1-x) + i \epsilon/p^2 \right)^2}
     \nonumber \\
     & ~+~
     \xi \Bigg[
          ~ \, \int^1_0dx  ~ \frac{x^2}{x(1-x) + i \epsilon/p^2} ~ + ~  \frac{1}{2}\, \int^1_0dx  ~ \frac{x^3}{\left(x(1-x) + i \epsilon/p^2 \right)^2}
     \Bigg]
      \Bigg\} 
 \ ,
\end{align}
The integration over Feynman parameters result in pure numbers, see App. \ref{App:FeynParameter}, and therefore for large photon or fermion momentum
$\tau_8$ scale with the inverse of the momentum squared. The perturbative estimation of this form factor, see \cite{Kizilersu:1995iz}, shares with
the above result the common  factor $\alpha / 4\, \pi \, \Delta$.

Let us proceed and look at the computation of $\tau_2$, $\tau_3$ and $\tau_6$. Within the approximations considered, 
their evaluation needs the quantity $\Lambda_1$ that is defined in Eq. (\ref{Eq:Lambda1Approx}). Following the same reasoning as in the previous cases, it comes
\begin{align}
\Lambda_1 & = 4 \, Z_2   
  ~ - ~ 4 \, i \, g^2 \,  \int \frac{d^4q}{(2 \, \pi)^4} ~ \frac{1}{q^2} 
  \frac{1}{(p-q)^2 - m^2} ~ \frac{1}{(p+k -q)^2 - m^2} 
  \nonumber \\
  & \hspace{3cm}
~  \Bigg\{  ~ p^2 + (pk)  - \frac{(pq)^2}{q^2} - \frac{(pq)(kq)}{q^2} ~  + ~  \xi \left[ q^2 - (kq) - 2 \, (pq)  + \frac{(pq)^2}{q^2} + \frac{(pq)(kq)}{q^2} \right]
 \Bigg\}  \ .
 \label{Eq:Lambda1ApproxPert0}
\end{align}
The momentum integration has a divergent term, the $\xi \, q^2$ term inside the integral. This divergence is exactly cancelled by the pole term of $Z_2$, see Eq. (\ref{Def:Z2Pert}), 
and $\Lambda_1$ is finite. Indeed, an evaluation of the divergent integral combined with the expression for $Z_2$ results in the finite expression
\begin{equation}
4 \, \Bigg[ 
1  ~ + ~ \frac{\alpha}{4 \, \pi} \, \xi \, \int^1_0dx \int^x_0dy ~ \Bigg\{
\frac{K^2}{\mathcal{M}^2 - K^2}  ~ + ~ 2  \, \ln \frac{\mathcal{M}^2 - K^2}{m^2} \Bigg\} 
~ - ~ \frac{\alpha}{4 \, \pi} \, \xi \,   \left( 1 + \frac{m^2}{\mu^2_F}  \right) \Big( 1 - L(\mu^2_F) \Big)
\Bigg]
\end{equation}
that should be added to
\begin{eqnarray}
&&
4 \, \frac{\alpha}{4 \, \pi} \,  \int^1_0dx\int^x_0dy \Bigg\{ 
  \frac{1 + \xi}{2} \, \frac{p^2 + (pk)}{\mathcal{M}^2 - K^2} 
                                                     ~ - ~ \xi ~\frac{ x \, \Big( (pk) + 2 \, p^2 \Big) ~ + ~ y \Big( k^2 + 2 \, (pk) \Big) }{\mathcal{M}^2 - K^2} 
                                                     \nonumber \\
&& \qquad\qquad
 + ~ (\xi -1)  
       ~ \frac{x^2 p^2 \Big( p^2+ (pk) \Big) \, + \, y^2 (pk) \Big( k^2 + (pk) \Big) \, + \, x \, y \, \Big( p^2 \big( k^2 + (pk) \big) \, + \, (pk) \big( p^2 + (pk) \big) \Big)}{\Big( \mathcal{M}^2 - K^2 \Big)^2} 
~ \Bigg\}  .
\end{eqnarray}
For small fermion momentum, it follows that
\begin{eqnarray}
\Lambda_1 & = & 
4 \, \Bigg\{  
1  ~ + ~ \frac{\alpha}{4 \, \pi} \, \xi \, \int^1_0dx \int^x_0dy ~ \Bigg(
        \frac{k^2}{m^2} \, \frac{y^2}{y(1-y) \frac{k^2}{m^2} -x + i \, \epsilon/m^2}  ~ + ~ 2  \, \ln \Big( y(1-y) \frac{k^2}{m^2} -x + i \, \epsilon/m^2 \Big)  \Bigg) 
        \nonumber \\
& & \qquad\quad
+ ~
\frac{\alpha}{4 \, \pi} \,  \int^1_0dx\int^x_0dy \Bigg(
   \frac{1 + \xi}{2} \, \frac{(pk)}{m^2}   ~ - ~ \xi ~\frac{k^2}{m^2} 
            + ~ (\xi -1)  ~ \frac{(pk) k^2}{m^4} \,  \frac{y }{ y(1-y) \frac{k^2}{m^2} -x + i \, \epsilon/m^2}  \Bigg) \times
            \nonumber \\
            & & \hspace{4.5cm}
              \times \frac{  y  }{y(1-y) \frac{k^2}{m^2} -x + i \, \epsilon/m^2}
        \nonumber \\
        & & \qquad\quad
~ - ~ \frac{\alpha}{4 \, \pi} \, \xi \,   \left( 1 + \frac{m^2}{\mu^2_F}  \right) \Big( 1 - L(\mu^2_F) \Big)
\Bigg\} 
\nonumber \\
& \longrightarrow &
4 \, \Bigg\{  
1  ~ + ~ \frac{\alpha}{4 \, \pi} \, \xi \,  \frac{k^2}{m^2} \ \int^1_0dx \int^x_0dy ~ 
        \frac{y (y -1) }{y(1-y) \frac{k^2}{m^2} -x + i \, \epsilon/m^2}   
\Bigg\} ~ \sim ~ 4 \, \left( 1  ~ + ~ \frac{\alpha}{8 \, \pi} \, \xi ~  \right) \ ,
\end{eqnarray}
where in the last line only the leading term in $k^2$ was taken into account.
The large photon momentum limit gives
\begin{eqnarray}
\Lambda_1 & = & 
4 \, \Bigg\{ 
1  ~ + ~ \frac{\alpha}{4 \, \pi} \, \xi \, \int^1_0dx \int^x_0dy ~ \Bigg(
\frac{y^2}{y(1-y) + i \epsilon/k^2}  ~ + ~ 2  \, \ln \left( \frac{k^2}{m^2}  \Big( y(1-y) + i \epsilon/k^2 \Big) \right) \Bigg)
\nonumber \\
 && \qquad\qquad
 - ~  4 \, \frac{\alpha}{4 \, \pi} \, \xi \,  \int^1_0dx\int^x_0dy ~    \frac{  y  }{y(1-y) + i \epsilon/k^2 } 
~ - ~ \frac{\alpha}{4 \, \pi} \, \xi \,   \left( 1 + \frac{m^2}{\mu^2_F}  \right) \Big( 1 - L(\mu^2_F) \Big)
~ \Bigg\}  
\nonumber \\
 & \longrightarrow &
 4 \, \Bigg\{ 
1  ~ + ~ \frac{\alpha}{4 \, \pi} \, \xi \,  \ln  \frac{k^2}{m^2} ~ \Bigg\}   \ .
\end{eqnarray}
At small fermion momentum
\begin{equation}
\lambda_1 = \frac{1}{2} \Big( A(k^2) + A(p^2) \Big) \ , \qquad
\lambda_2 = \frac{1}{2 \, k^2} \Big( A(k^2) - A(p^2) \Big) \ , 
\end{equation}
and, therefore,
\begin{eqnarray}
\tau_2 & = & - \, \frac{1}{\Delta}  \Bigg(  1  \, + \, \frac{\alpha}{8 \, \pi} \, \xi  \, - \, \frac{3}{8} \Big( A(k^2) + A(p^2) \Big) \, - \, \frac{3}{4} \, \frac{(pk)}{k^2} \, \Big( A(k^2) - A(p^2) \Big) \Bigg)
\\
\tau_3 & = & \frac{k^2}{2 \, \Delta} \Bigg( 1  \, + \, \frac{\alpha}{8 \, \pi} \, \xi  \,  + \, \frac{3}{4} \Big( A(k^2) + A(p^2) \Big) \, + \, \frac{1}{4} \, \frac{(pk)}{k^2}  \, \Big( A(k^2) - A(p^2) \Big) \Bigg)
\\
\tau_6 & = & - \, \tau_3 \ ,
\end{eqnarray}
while at large photon momentum the form factors are given by the same expressions with the substitution
\begin{equation}
1  \, + \, \frac{\alpha}{8 \, \pi} \, \xi  \quad \longrightarrow \quad 1  ~ + ~ \frac{\alpha}{4 \, \pi} \, \xi \,  \ln  \frac{k^2}{m^2} \ .
\end{equation}
For the Landau gauge where $A(k^2) = 1$ and $\xi = 0$, the above expressions simplify into
\begin{eqnarray}
\tau_2  =  - \, \frac{1}{4 \, \Delta}  \ , \qquad\mbox{ and }\qquad \tau_3  =  \frac{5 \, k^2}{4 \, \Delta} = \frac{5}{4 \, \Big( p^2 - \frac{(pk)^2}{k^2} \Big)}\ .
\end{eqnarray}
In both limits considered and within the approximation discussed, $\tau_3 + \tau_6 = 0$ and the full vertex, see Eq. (\ref{FullVertex-Dirac-main}), simplifies into
\begin{eqnarray}
{\Gamma}^\mu (p, \, -p -k; \, k)   & = &
     {p}^{\mu } ~  \Big(\, {k}^2 \,  {\tau_1} \,  - \, 2 \, {\lambda_3} \, \Big) - {k}^{\mu }  ~ \Big(\, (pk) \,  {\tau_1} \,  + \, {\lambda_3} \Big)
     \nonumber \\
     & & 
     + \, 
     {\gamma }^{\mu } \Bigg[ \, {\lambda_1} \, - \, \, 2 \,  ( {p} {k}) \, {\tau_3} \, \Bigg]
     + \, \slashed{k}  \, \Bigg[ \, {k}^{\mu } \Big( \, {\lambda_2} \, - \,  ({p} {k} ) \,  {\tau_2} \Big) 
                                             \, + \, {p}^{\mu } \Big(2 \, {\lambda_2} \, + \, k^2 \, \, {\tau_2} \, + \, 2 \, {\tau_3} \Big) \, \Bigg]
     \nonumber \\
     & & \hspace{3.75cm}
     + \, 2 \, \slashed{p} \, \Bigg[ \, {k}^{\mu } \Big(\, {\lambda_2} \, -  \, ({k} {p}) \, {\tau_2}\Big) \, + \, {p}^{\mu } \Big( 2 \, {\lambda_2} \, + \, k^2 \, {\tau_2} \Big)\,  \Bigg]
     \nonumber \\
     & & 
     + \, i ~~ {\tau_8}  ~~ {\gamma }_{\sigma } ~ {\gamma }_5 ~~ ~{\epsilon }^{\sigma  \mu  \alpha\beta} ~~  {p}_\alpha \, {k}_\beta
     \ .
     \label{FullVertex-Dirac-main-App-Pert}
\end{eqnarray}
In the Landau gauge, where $\xi = 0$ and $A(k^2) = 1$, the above expressions can be simplified further.

\section{Summary and Conclusions \label{Sec:Summary}}

In the current work the DSE for the propagators together with the photon-fermion vertex are investigated. After the introduction of
a tensor basis to describe the vertex, a set of exact nonlinear integral equations for each of the transverse form factors is derived.
These equations are quite general and can be applied both to the QED photon-fermion vertex and to the QCD quark-gluon vertex, after taking into account the 
color structure and correcting for the contribution of $\lambda_4$. 
The nonlinear integral equations allow, in principle, to compute the full vertex, assuming that the longitudinal part of the vertex is given by the solution of
the vertex Ward-Takahashi identity. Indeed, as reported, the equations are solved, for any linear covariant gauge, within a perturbative approach.
The comparison of the solutions of the integral equation with the result of one-loop perturbation theory show that the main features are shared by the two solutions of QED.
For example,
in the chiral limit both solutions coincide in the number of form factors that are needed to describe fully $\Gamma^\mu$.
Being a first step to access the full vertex, this result is encouraging and motivates the realisation of a non-perturbative study of the vertex in QED. 
Non-perturbative studies can be performed e.g. by going to the Euclidean spacetime or by calling for integral representations of the form factors. 

Any solution of the DSE requires defining a truncation of the infinite tower of relations. For QED, assuming that perturbation theory makes sense,
there is a ``natural'' way of approaching the infinite tower of integral equations. Herein, the truncation considered ignores the contribution of the 
two-photon-two-fermion one-particle irreducible Green function. This drastic approximation can be improved as it can be found in the literature a solution
of the corresponding WTI, that can combined with the corresponding DSE. This integral equation is also a function of the full vertex, making the realisation of
such an improvement not trivial. What changes by improving the solution of the DSE for e.g. chiral symmetry breaking and the
number of critical fermion flavours \cite{Bashir:2011ij} remains to be seen. Using the information that can be read from the vertex integral equation can also help to improve
the modelling the photon-fermion vertex and its compatibility with the requirements of QFT.

A full study of the system of equations requires also a proper handling of the infrared divergences, that were briefly touched when looking at the photon propagator equation.
This issue is also important in the definition of an on-shell vertex and its relation to the fermion anomalous magnetic and electric moments. In particular, 
in the current manuscript the on-shell vertex is given in terms of the fundamental form factors that are associated with $\Gamma^\mu$.

\section*{Acknowledgements}

This work was financed through national funds by FCT - Funda\c{c}\~ao para a Ci\^encia e a Tecnologia, I.P. in the framework of the projects UIDB/04564/2020 and UIDP/04564/2020, with DOI identifiers 10.54499/UIDB/04564/2020 and 10.54499/UIDP/04564/2020, respectively.
The author also acknowledges financial support from  grant 2022/05328-3, from São Paulo Research Foundation (FAPESP).
The author thanks T. Frederico and W. de Paula for helpful discussions.

\begin{appendix}

\section{Dyson-Schwinger Equations with full basis of operators \label{AppDSEFull}}

For completeness, in this section, we report on the DSE using the full basis discussed
in the main text. The Dirac algebra was worked out with the help FeynCalc \cite{FeynCalc1,FeynCalc2,FeynCalc3},
a \textit{Mathematica} software package.

\subsection{The fermion gap equation \label{Sec:PropEqGap}}

The fermion gap equation (\ref{DSE-Fermion}) can be projected into a Dirac scalar and a Dirac vector term. 
The scalar component of the equation is obtained by taking the Dirac trace of Eq. (\ref{DSE-Fermion}) that, in a general linear covariant
gauge, reads
\begin{eqnarray}
& & 
  B (p^2)  ~ = ~  m 
     - \, i \, g^2 \, \int \frac{d^4 k}{( 2 \, \pi )^4} ~  \frac{1}{A^2((p-k)^2) \,(p-k)^2 - B^2((p-k)^2)}  \Bigg\{
     \nonumber \\
     & & \hspace{2cm}
D(k^2)
 \Bigg( ~~
     A((p-k)^2) 
      \Bigg[ ~ 
      2 \, \lambda_3 \, p^2 \bigg( \frac{(pk)^2}{p^2 \, k^2} -  1\bigg)
              ~ + ~ \tau_1 \, k^2 \, p^2 \bigg( 1 - \frac{(pk)^2}{k^2 \, p^2} \bigg)
      \nonumber \\
      & & \hspace{5.3cm}\qquad
              + ~ \tau_4 \, k^4 \, p^2 \bigg( 1 -  \frac{ (pk) }{k^2} -  \frac{(pk)^2}{p^2 \, k^2}  + \frac{(pk)^3}{p^2 \, k^4}  \bigg)
              ~ + ~ 3 \, \tau_5 \, k^2 \bigg( 1 - \frac{(pk)}{k^2} \bigg)
      \nonumber \\
      & & \hspace{5.3cm}\qquad
           + ~   \tau_7 \, k^4 \bigg( \frac{3 }{2}  + 4 \,  \frac{p^2}{k^2}  - \frac{15 }{2} \, \frac{ (pk) }{k^2} - 6 \, \frac{p^2}{k^2} \, \frac{(pk)}{k^2} + 8 \,  \frac{(pk)^2}{k^4} \bigg)
      \Bigg]
      \nonumber \\
      & & \hspace{3.5cm}
     + ~ B((p-k)^2)  
           \Bigg[
              3 \, \lambda_1
              ~ + ~ 4 \, \lambda_2 \, p^2 \bigg( 1 - \frac{ (pk)^2}{p^2 \, k^2} \bigg)
              ~ + ~ 2 \, \tau_2\, k^2 \, p^2 \bigg( 1 -  \frac{(pk)^2}{p^2 \, k^2} \bigg)
              \nonumber \\
              & & \hspace{5.5cm}\qquad
              ~ + ~ 3 \,  \tau_3 \,  k^2
              + ~ 3 \, \tau_6 \, k^2 \bigg( 2 \, \frac{(pk)}{k^2} -  1 \bigg)
            \Bigg] ~~
\Bigg)
\nonumber \\
& & \hspace{2.5cm}
+ ~ \frac{\xi}{k^2} 
    \Bigg( ~
       A((p-k)^2)  \Bigg[   \lambda_3 \, k^2 \bigg( -\frac{2 (pk)^2}{k^4} + 3 \, \frac{(pk)}{k^2} - 1 \bigg)
                         \Bigg]
       \nonumber \\
       & & \hspace{5.5cm}\qquad
       + ~ B((p-k)^2) \Bigg[  \lambda_1 ~ + ~ \lambda_2 \, k^2 \left(\frac{4 (pk)^2}{k^4} - 4 \, \frac{(pk)}{k^2}  + 1\right)  \Bigg] ~
    \Bigg)
    ~\Bigg\}     \ .
      \label{DSE-Fermion-scalar}
\end{eqnarray}
The vector component is obtained after multiplying Eq. (\ref{DSE-Fermion}) by $\slashed{p}$ and then taking the Dirac trace 
\begin{eqnarray}
& & 
 A(p^2)  = 1
     + \, i \, g^2 \, \int \frac{d^4 k}{( 2 \, \pi )^4} ~   \frac{1}{A^2((p-k)^2) \,(p-k)^2 - B^2((p-k)^2)}\Bigg\{
\nonumber \\
& & \qquad
D(k^2) 
 \Bigg(
     A((p-k)^2) \Bigg[ ~
         \lambda_1 \,  \bigg( - 1 + 3 \, \frac{(pk)}{p^2} - \, 2 \, \frac{(pk)^2}{p^2 \, k^2}\bigg)
        \nonumber \\
        & & \hspace{3.5cm}\qquad
        + ~ 2 \, \lambda_2 \, k^2  \bigg( 1 + 2 \, \frac{p^2}{k^2}  - 2 \, \frac{ (pk) }{k^2} -  \frac{(pk)^2}{p^2 \, k^2}
                                                                      - \, 2 \,  \frac{ (pk)^2}{k^4} + 2 \, \frac{(pk)^3}{p^2 \, k^4} \bigg)
         \nonumber \\
        & &     \hspace{3.5cm}\qquad
         + ~ \tau_2 \, k^4  \bigg( 1 + 2  \, \frac{p^2}{k^2}  - 2 \,  \frac{(pk)}{k^2}  - \frac{(pk)^2}{p^2 \, k^2}  - 2 \,  \frac{ (pk)^2}{k^4} + 2  \, \frac{(pk)^3}{p^2 \, k^4} \bigg)
        \nonumber \\
        & & \hspace{3.5cm} \qquad
        + ~ \tau_3 \, k^2  \bigg( - \, 1 + 3 \, \frac{(pk)}{p^2} - 2 \, \frac{(pk)^2}{p^2 \, k^2} \bigg)
        \nonumber \\
        & & \hspace{3.5cm}  \qquad
        + ~ 3 \, \tau_6 \, k^2 \bigg(  1 -  \frac{(pk)}{p^2} - 2 \frac{(pk)}{k^2} + 2 \frac{(pk)^2}{p^2 \, k^2} \bigg)
        ~ + ~ 2\, \tau_8 \, k^2  \bigg( 1 -  \frac{(pk)^2}{p^2 \, k^2} \bigg)
                     \Bigg]
     \nonumber \\
     & &  \hspace{1.5cm}\qquad
     +~ B((p-k)^2)  \Bigg[~
                              2 \, \lambda_3 \,  \bigg(\frac{(pk)^2}{p^2 \, k^2} -  1 \bigg)
                             ~ + ~ \tau_1 \, k^2  \bigg( 1 - \frac{(pk)^2}{p^2 \, k^2} \bigg)
                             ~ + ~ \tau_4 \, k^2  (pk) \, \bigg( 1 -  \frac{(pk)^2}{p^2 \, k^2} \bigg)
        \nonumber \\
        & & \hspace{4.5cm}\qquad        
                             + ~ 3 \, \tau_5 \, \frac{(pk)}{p^2}
                             ~ + ~\tau_7 \, k^2 \,  \bigg(- \, 2  + \frac{3}{2} \,  \frac{(pk)}{p^2}  + 6 \, \frac{(pk)}{k^2}  - 4 \frac{(pk)^2}{p^2 \, k^2} \bigg)
                             \Bigg]
     \Bigg)
 \nonumber \\
 & & \qquad
 + ~ \frac{\xi}{k^2}
   \Bigg(
   A(p-k)^2)  \Bigg[
               \lambda_1 \,  \bigg( - 1 \, - \frac{(pk)}{p^2}  + 2 \, \frac{ (pk)^2}{p^2 \, k^2} \bigg)
               \nonumber \\
               & & \hspace{3.5cm}\qquad
               + ~ \lambda_2 \, k^2 \bigg( 1 - \frac{(pk)}{p^2} - 4 \, \frac{(pk)}{k^2}  + 4 \,  \frac{(pk)^2}{p^2 \, k^2}  + 4 \, \frac{(pk)^2}{k^4} - 4 \, \frac{(pk)^3}{p^2 \, k^4} \bigg)
               \Bigg]
   \nonumber \\
   & & \hspace{1.5cm}\qquad
   + ~ B(p-k)^2) \Bigg[ \lambda_3 \,  \bigg( \frac{(pk)}{p^2} - 2 \, \frac{(pk)^2}{p^2 \, k^2}\bigg)
                     \Bigg] ~
   \Bigg)
~ \Bigg\} \ ,
     \label{DSE-Fermion-vector}
\end{eqnarray}
where in these equations
\begin{equation}
\lambda_i = \lambda_i (p^2, \, (p-k)^2,\, k^2 ) \qquad\mbox{ and }\qquad
\tau_i = \tau_i ( p^2, \, (p-k)^2,\, k^2 )  \ .
\label{TransFF-GapEq}
\end{equation}
The fermion self-energies $\Sigma_s$ and $\Sigma_v$, see Eq. (\ref{DSER-gap}) for the definition,
can be read from Eqs (\ref{DSE-Fermion-scalar}) and (\ref{DSE-Fermion-vector}).
This decomposition of the gap equation in terms of vertex form factors holds both for QED and for QCD, after
correcting for the color degrees of freedom and adding the contribution of $\lambda_4$, that vanishes in QED.

\subsection{The photon gap equation \label{Sec:PropEqPhoton}}

In the basis considered, the equation for the photon propagator is given by
\begin{eqnarray}
& &
  \frac{1}{D(k^2)}  = 
   k^2
  - i \, \frac{g^2}{3} \,  \int \frac{d^4 p}{(2 \, \pi)^4} ~  \frac{1}{A^2\left(\left(p + \frac{k}{2}\right)^2\right) \, (p + \frac{k}{2})^2 - B^2\left(\left(p + \frac{k}{2}\right)^2\right)} 
  \nonumber \\
  & & \hspace{4.5cm}
                                                                             ~ \frac{1}{A^2\left(\left(p - \frac{k}{2}\right)^2\right) \, (p-\frac{k}{2})^2 - B^2\left(\left(p - \frac{k}{2}\right)^2\right)}  \nonumber \\
  & & 
   \Bigg\{ ~
A\left(\left(p + \frac{k}{2}\right)^2\right) A\left(\left(p - \frac{k}{2}\right)^2\right) 
\Bigg[
 2 \, \lambda_1 \,  \bigg( 1 - 4 \, \frac{p^2}{k^2} \bigg)
 ~ + ~  4 \, \lambda_2   \bigg(  \frac{ 4 \, p^4 - 2 \, (pk)^2}{ k^2} + p^2 \bigg)
 \nonumber \\
 & & \hspace{6cm}  
 + ~  2 \, \tau_2  \,  \bigg(  - 4 \, \frac{p^2  (pk)^2}{k^2} + k^2 p^2 + 4 \, p^4 - (pk)^2 \bigg)
   \nonumber \\
   & & \hspace{6cm}                    
   + ~  \, \tau_3 \,  \bigg( - 8 \, \frac{(pk)^2}{k^2} + 3 \, k^2 - 4 \, p^2 \bigg)
   \nonumber \\
   & & \hspace{6cm}                    
   + ~ 6 \, \tau_6 \,  \bigg(  4 \, p^2 \, \frac{(pk) }{k^2} - \, (pk) \bigg)
   ~ + ~8 \,  \tau_8 \, \bigg(  p^2 - \frac{(pk)^2}{k^2}  \bigg)
\Bigg]
\nonumber \\
& &
\quad + ~  A\left(\left(p + \frac{k}{2}\right)^2\right) B\left(\left(p - \frac{k}{2}\right)^2\right) 
\Bigg[ 
   \, - \, 4 \,  \lambda_3 \, \bigg( \frac{ (pk)}{k^2} + 2 \, \frac{ p^2}{k^2}\bigg) 
   ~ + ~ 4 \, \tau_1 \,   \bigg(  p^2  - \frac{ (pk)^2}{k^2} \bigg)
   \nonumber \\
   & & \hspace{6cm}
  ~ + ~  2\, \tau_4 \,  \bigg(  p^2 k^2  + 2 \, p^2  (pk)   - 2 \, (pk)^2 - 2 \, \frac{(pk)^3}{k^2} \bigg)
~ + ~ 6 \, \tau_5 \,\bigg( 1 +  2 \, \frac{(pk)}{k^2}  \bigg)
   \nonumber \\
   & & \hspace{6cm}                    
   ~ + ~  4 \, \tau_7 \,  \bigg(   p^2 + 2 \, \frac{(pk)^2}{k^2} + 6 \, p^2 \, \frac{(pk)}{k^2}  \bigg)
\Bigg]
\nonumber \\
& &
\quad + ~  B\left(\left(p + \frac{k}{2}\right)^2\right) A\left(\left(p - \frac{k}{2}\right)^2\right)
\Bigg[
   4 \, \lambda_3 \,  \bigg(  \frac{  (pk) }{k^2} - 2 \, \frac{ p^2}{k^2} \bigg)
   ~ + ~ 4 \, \tau_1 \,  \bigg( p^2 - \frac{(pk)^2}{k^2}  \bigg)
 \nonumber \\
 & & \hspace{6cm}               
 + ~ 2  \, \tau_4 \, \bigg(  p^2 k^2   - 2 \, p^2 (pk) -  2 \, (pk)^2 + 2 \, \frac{(pk)^3}{k^2} \bigg)
 ~ + ~ 6 \, \tau_5 \,  \bigg( 1 - 2 \, \frac{(pk)}{k^2} \bigg)
 \nonumber \\
 & & \hspace{6cm}                 
 + ~ 4 \, \tau_7 \, \bigg(   p^2 + 2 \, \frac{(pk)^2 }{k^2} - 6 \, p^2 \, \frac{ (pk)}{k^2}   \bigg)
\Bigg]
\nonumber \\
& &
+B\left(\left(p + \frac{k}{2}\right)^2\right) B\left(\left(p - \frac{k}{2}\right)^2\right) 
\Bigg[
16 \, \lambda_1 \frac{1}{k^2} 
~ +~ 16 \, \lambda_2 \,   \frac{p^2}{k^2}
~ + ~ 8 \, \tau_2 \,  \bigg(  p^2 - \frac{(pk)^2}{k^2} \bigg)
 ~ + ~ 12 \, \tau_3 \, 
 ~ - ~ 24 \, \tau_6 \,  \frac{(pk)}{k^2} 
\Bigg] ~
\Bigg\} \ ,
     \label{Eq:PhotonEqBasis}
\end{eqnarray}
where the form factors are defined by
\begin{equation}
\lambda_i = \lambda_i \big( (p-k/2)^2, \, (p + k/2)^2, \, k^2\big)
\qquad\mbox{ and }\qquad 
\tau_i = \tau_i \big( (p-k/2)^2, \, (p + k/2)^2, \, k^2\big) \ .
\label{TransFF-PhotonGapEq}
\end{equation}
As for the fermion gap equation, the photon DSE gets contributions from all the transverse form factors.

\subsection{The photon-fermion vertex equation \label{SecVertex}}

The decomposition of the photon-fermion vertex in terms of Dirac bilinear forms within the basis defined by Eqs (\ref{TensorBasis-Long}) to (\ref{TensorBasis-Ortho})
is given by
\begin{eqnarray}
& & 
{\Gamma}^\mu (p, \, -p -k; \, k)   ~ = ~ 
     {p}^{\mu } ~  \Big(\, {k}^2 \,  {\tau_1} \,  - \, 2 \, {\lambda_3} \, \Big) - {k}^{\mu }  ~ \Big(\, (pk) \,  {\tau_1} \,  + \, {\lambda_3} \Big)
     \hspace{1.5cm}\text{(Scalar Component)}
     \nonumber \\
     & & 
     \qquad\qquad
     + \, 
     {\gamma }^{\mu } \Bigg[ \, {\lambda_1} \, + \, k^2 \, {\tau_3}+\,  \Big( k^2  \, + \, \, 2 \,  ( {p} {k}) \Big) \, {\tau_6} \, \Bigg]
      \hspace{3.6cm}\text{(Vector Components)}
     \nonumber \\
     & & \qquad\qquad\qquad
     + \, \slashed{k}  \, \Bigg[ \, {k}^{\mu } \Big( \, {\lambda_2} \, - \,  ({p} {k} ) \,  {\tau_2} \, - \, {\tau_3} \, -\, {\tau_6}\Big) 
                                             \, + \, {p}^{\mu } \Big(2 \, {\lambda_2} \, + \, k^2 \, \, {\tau_2}- \, 2 \, {\tau_6} \Big) \, \Bigg]
     \nonumber \\
     & & \qquad\qquad\qquad
     + \, 2 \, \slashed{p} \, \Bigg[ \, {k}^{\mu } \Big(\, {\lambda_2} \, -  \, ({k} {p}) \, {\tau_2}\Big) \, + \, {p}^{\mu } \Big( 2 \, {\lambda_2} \, + \, k^2 \, {\tau_2} \Big)\,  \Bigg]
     \nonumber \\
     & & \qquad\qquad
     + \, i ~~ {\tau_8}  ~~ {\gamma }_{\sigma } ~ {\gamma }_5 ~~ ~{\epsilon }^{\sigma  \mu  \alpha\beta} ~~  {p}_\alpha \, {k}_\beta
           \hspace{5.3cm}\text{(Axial Component)}
     \nonumber \\
     & & \qquad\qquad
     + \,  
     \frac{1}{2} \, \sigma^{\mu\alpha} \, k_\alpha \,  \Bigg[ 2 \, \tau_5 \, + \, \tau_7 \Big( k^2 + 2 \, (kp) \Big) \Bigg]
     \hspace{4.cm}\text{(Tensor Components)}
     \nonumber \\
     & & \qquad\qquad\qquad
     + \,  
          \sigma^{\mu\alpha} \, p_\alpha \,  \Bigg[ \,  \tau_7 \Big( k^2 + 2 \, (kp) \Big) \, \Bigg]
     \nonumber \\
     & & \qquad\qquad\qquad
     \, + \,  \sigma^{\alpha\beta} \, p_\alpha \, k_\beta \Bigg[ \, k^\mu \, \Big( \tau_7 -  (pk) \, \tau_4  \Big) +  p^\mu \Big(  k^2 \, \tau_4  +  2 \, \tau_7  \Big) \Bigg]
     \label{FullVertex-Dirac}
\end{eqnarray}
with the form factors being
\begin{equation}
   \lambda_i = \lambda_i \Big((p+k)^2, \, p^2, \, k^2\Big)
   \qquad\mbox{ and }\qquad
   \tau_i = \tau_i \Big((p+k)^2, \, p^2, \, k^2\Big) \ .
\end{equation}   
The above decomposition is useful to rewrite the DSE for the vertex.

\section{Momentum Integration \label{App:MomInt}}

The integrations over momentum that appear in the evaluation of the perturbative like solution of the vertex equation can be
performed, within dimensional regularisation, with the help of Feynman parametrisation. Indeed, one can write
\begin{equation}
\frac{1}{q^2} ~\frac{1}{ (p-q)^2 - m^2}~ \frac{1}{(p + k  -q)^2 - m^2}  ~ =  ~
\Gamma (3) \, \int^1_0 dx  \int^x_0 dy ~
 \frac{1}{\Big(q^2 + 2 (Kq) +  \mathcal{M}^2\Big)^3} \ ,
 \end{equation}
together with
\begin{equation}
\frac{1}{q^4} ~\frac{1}{ (p-q)^2 - m^2}~ \frac{1}{(p + k  -q)^2 - m^2}  ~  = ~ 
\Gamma(4) \, \int^1_0 dx  \int^x_0 dy ~
\frac{y}{\Big(q^2 + 2 (Kq) +  \mathcal{M}^2\Big)^4}
\end{equation}
where $K = -x \, p - y k$ and  $\mathcal{M}^2 = x\, ( p^2 - m^2) + y ( k^2 + 2 \, (pk) )$. It follows that the integrations over
$q$ to consider are
\begin{eqnarray} 
I_0 & = &
    \int \frac{d^nq}{(2 \, \pi)^n} ~ \frac{1}{q^2} ~\frac{1}{ (p-q)^2 - m^2}~ \frac{1}{(p + k  -q)^2 - m^2} 
    \nonumber \\
    &  = &  \frac{i}{16 \, \pi^2} ~ \int^1_0 dx \int^x_0 dy ~ \frac{1}{ \mathcal{M}^2 - K^2}  
                   \left[ 1 + \frac{\epsilon}{2} \left( \ln \frac{\mathcal{M}^2 - K^2}{\zeta^2}  + \ln 4 \, \pi - \gamma_E \right)  + \cdots \right] \ ,\\
I_\mu & = &
    \int \frac{d^nq}{(2 \, \pi)^n} ~ q_\mu ~\frac{1}{q^2} ~\frac{1}{ (p-q)^2 - m^2}~ \frac{1}{(p + k  -q)^2 - m^2} 
    \nonumber \\
    &  = & \frac{i}{16 \, \pi^2} ~ \int^1_0 dx \int^x_0 dy ~ \frac{~~ x \, p_\mu + y \, k_\mu ~~}{ \mathcal{M}^2 - K^2} 
                 \left[ 1 + \frac{\epsilon}{2} \left( \ln \frac{\mathcal{M}^2 - K^2}{\zeta^2}  + \ln 4 \, \pi - \gamma_E \right)  + \cdots \right] \ , \\
I_{\mu\nu} & = &
    \int \frac{d^nq}{(2 \, \pi)^n} ~ q_\mu q_\nu ~\frac{1}{q^2} ~\frac{1}{ (p-q)^2 - m^2}~ \frac{1}{(p + k  -q)^2 - m^2} 
    \nonumber \\
    &  = & \frac{i}{16 \, \pi^2} ~ \int^1_0 dx \int^x_0 dy ~ \frac{1}{ \mathcal{M}^2 - K^2} 
              \bigg(  \big( x \, p_\mu + y \, k_\mu \big) \big( x \, p_\nu + y \, k_\nu \big)  \big( 1 + \cdots \big)
                       \nonumber \\
                       & & \hspace{5.3cm}
                           + \frac{1}{2} g_{\mu\nu} \left( \mathcal{M}^2 - K^2 \right) 
                 \left[ \frac{2}{\epsilon} + \ln \frac{\mathcal{M}^2 - K^2}{\zeta^2}  + \ln 4 \, \pi - \gamma_E + \cdots \cdots \right]  \bigg)\ ,
\end{eqnarray}
where it was used $n = 4 - \epsilon$,  $\zeta$ is usual mass introduced in the dimensional regularization and $\gamma_E$ is the Euler-Mascheroni constant.
We call the reader attention to the change of notation in this App. and that used to write the results in Eqs (\ref{Eq:PertA}) and (\ref{Eq:PertB}) and subsequent 
results that follow the notation of \cite{Kizilersu:1995iz}. Besides the above integrals, the perturbative solution of the vertex equation also needs
the following finite integrals
\begin{eqnarray} 
\widetilde{I}_0 & = &
    \int \frac{d^4q}{(2 \, \pi)^4} ~ \frac{1}{q^4} ~\frac{1}{ (p-q)^2 - m^2}~ \frac{1}{(p + k  -q)^2 - m^2} 
        \nonumber \\
         &  = & \frac{i}{16 \, \pi^2} ~ \int^1_0 dx \int^x_0 dy ~ \frac{1}{ \left( \mathcal{M}^2 - K^2 \right)^2}  \ , \\
\widetilde{I}_\mu & = &
    \int \frac{d^4q}{(2 \, \pi)^4} ~ q_\mu ~\frac{1}{q^4} ~\frac{1}{ (p-q)^2 - m^2}~ \frac{1}{(p + k  -q)^2 - m^2} 
        \nonumber \\
        &   =  &  \frac{i}{16 \, \pi^2} ~ \int^1_0 dx \int^x_0 dy ~ \frac{~~ x \, p_\mu + y \, k_\mu ~~}{ \left( \mathcal{M}^2 - K^2 \right)^2}  \ ,\\
\widetilde{I}_{\mu\nu} & = &
    \int \frac{d^4q}{(2 \, \pi)^4} ~ q_\mu ~\frac{1}{q^4} ~\frac{1}{ (p-q)^2 - m^2}~ \frac{1}{(p + k  -q)^2 - m^2} \nonumber \\
    &  = & \frac{i}{16 \, \pi^2} ~ \int^1_0 dx \int^x_0 dy ~ \frac{1}{ \left( \mathcal{M}^2 - K^2 \right)^2} 
           \Bigg( (x \, p_\mu + y \, k_\mu) (x \, p_\nu + y \, k_\nu) + \frac{g_{\mu\nu}}{2} \left( \mathcal{M}^2 - K^2 \right) \Bigg) \ , \\
\widetilde{I}_{\mu\nu\eta} & = &
    \int \frac{d^4q}{(2 \, \pi)^4} ~ q_\mu ~\frac{1}{q^4} ~\frac{1}{ (p-q)^2 - m^2}~ \frac{1}{(p + k  -q)^2 - m^2} \nonumber \\
    &  = & \frac{i}{16 \, \pi^2} ~ \int^1_0 dx \int^x_0 dy ~ \frac{1}{ \left( \mathcal{M}^2 - K^2 \right)^2} 
             \nonumber \\
             & & \qquad
           \Bigg[ (x \, p_\mu + y \, k_\mu) (x \, p_\nu + y \, k_\nu)(x \, p_\eta + y \, k_\eta) 
                    \nonumber \\
                    & & \qquad\qquad\qquad
                    + ~
                    \frac{1}{2} \bigg( g_{\mu\nu} (x \, p_\eta + y \, k_\eta ) 
                                               + g_{\mu\eta} (x \, p_\nu + y \, k_\nu )  
                                               + g_{\nu\eta} (x \, p_\mu + y \, k_\mu ) \bigg) \left( \mathcal{M}^2 - K^2 \right) \Bigg] \ .
\end{eqnarray}

\subsection{Integration over Feynman parameters \label{App:FeynParameter}}

After performing the momentum integrations and in the asymptotic regime one is left with the integration over Feynman parameters of the type
\begin{equation}
I(n,a) = \int^a_0dx ~\frac{x^n}{x(1-x) + i \epsilon/k^2} \ ,
\end{equation}
In order to perform the integration we rewrite the denominator as
\begin{equation}
\text{Den} = - ( x - z_0) (x -z_1) \ ,
\end{equation}
where
\begin{equation}
z_0 = \frac{1 + e^{ i \, \eta \, \pi/2}}{2} = \frac{1}{2} \left( 1 + i \, \sin \left( \frac{\eta \, \pi}{2} \right) \right) \ , \qquad\qquad z_1 = \frac{1 - e^{ i \, \eta \, \pi/2}}{2}
= \frac{1}{2} \left( 1 - i \, \sin \left( \frac{\eta \, \pi}{2} \right) \right) 
\end{equation}
and
\begin{equation}
\eta = \left\{ \begin{array}{lll}
                 +1, & & \text{for } ~ k^2 > 0 \ , \\
                 \\
                 -1, & & \text{for }  ~ k^2 < 0  \ .
                 \end{array} \right.
\end{equation} 
With these definitions 
\begin{equation}
I(n,a) = - e^{ - i \, \eta \, \pi/2} \int^a_0dx ~ x^n \left( \frac{1}{x - z_0} - \frac{1}{x-z_1} \right) \ .
\end{equation}
Then, the integrals read
\begin{eqnarray}
I(0,a) & = & i \,  \left( \ln \frac{a - z_0}{a-z_1} + \ln \frac{z_1}{z_0} \right) \ , \\
I(1,a) & = & i \,  \left( z_0  \, \ln \frac{ z_0 - a}{z_0} - z_1 \, \ln \frac{z_1-a}{z_1}  \right) \ , \\
I(2,a) & = & i \,  \left( a \big( z_0 - z_1\big) + z^2_0 \, \ln \frac{z_0 -a}{z_0} - z^2_1 \, \ln \frac{z_1 -a}{z_1} \right) \ , \\
I(3,a) & = & i \,  \left( \frac{a^2}{2} \big( z_0 - z_1\big) + a^3 \big( z^2_0 - z^2_1\big) + z^3_0 \, \ln \frac{z_0 -a}{z_0} - z^3_1 \, \ln \frac{z_1 -a}{z_1} \right) 
\end{eqnarray}
and they are pure complex numbers.

\end{appendix}
 

\end{document}